\documentclass[10pt,aps,prb,showpacs,twocolumn,amsmath,amssymb,superscriptaddress]{revtex4-1}

\usepackage{bbm}
\usepackage{mathtools}
\usepackage{graphicx}
\usepackage{multirow}

\begin{document}

\title{Boundary scattering effects on magnetotransport of narrow metallic wires and films}
\author{E.\ Khalaf}
\affiliation{Max Planck Institute for Solid State Research, Heisenbergstr.\ 1, 70569 Stuttgart, Germany}
\author{P.\ M.\ Ostrovsky}
\affiliation{Max Planck Institute for Solid State Research, Heisenbergstr.\ 1, 70569 Stuttgart, Germany}
\affiliation{L.\ D.\ Landau Institute for Theoretical Physics, 142432 Chernogolovka, Russia}

\begin{abstract}
Electron transport in thin metallic wires and films is strongly influenced by the quality of their surface. Weak localization and magnetoconductivity are also sensitive to the electron scattering at the edges of the sample. We study weak localization effects in a two-dimensional electron gas patterned in the form of a narrow quasi-one-dimensional channel in transverse magnetic field. The most general boundary conditions interpolating between the limits of mirror and diffuse edge scattering are assumed. We calculate magnetoconductivity for an arbitrary width of the sample including the cases of diffusive and ballistic lateral transport as well as the crossover between them. We find that in a broad range of parameters, the electron mobility is limited by the boundary roughness while the magnetotransport is only weakly influenced by the quality of the edges. In addition, we calculate magnetoconductivity for a metallic cylinder in the transverse field and a quasi-two-dimensional metallic film in the parallel field. 
\end{abstract}

\pacs{
73.63.-b, 	%Electronic transport in nanoscale materials and structures
72.15.Rn, 	%Localization effects (Anderson or weak localization)
%Electronic transport phenomena in thin films
73.50.Jt, 	%Galvanomagnetic and other magnetotransport effects (including thermomagnetic effects)
73.50.Dn 	%Low-field transport and mobility; piezoresistance
}

\maketitle

\section{Introduction}

Quantum interference is crucial to the transport properties of metals at low temperatures (for a review see Ref.\ \onlinecite{LeeRamakrishnan}). The most experimentally relevant manifestation of the quantum nature of transport is the low-field magnetoresistance. This effect is strongest in samples with restricted geometry, especially, thin films and wires. The origin of magnetoresistance is the interference between different trajectories of coherent electron propagation. Such interference is particularly strong for self-intersecting trajectories, where an electron can traverse a closed loop in either of two possible directions. When the interference is constructive, it enhances the probability of return to the same point and, thus, increases the resistivity. This effect is known as weak localization. External magnetic field, piercing the trajectory loop, destroys interference and decreases the resistivity, leading to negative magnetoresistance. In the case of destructive electron interference, that usually happens in materials with strong spin-orbit coupling, the sign of the effect is opposite, hence, the name of the effect is weak antilocalization and it leads to positive magnetoresistance.

The theory of weak localization was developed by Gor'kov et al.\ in Ref.\ \onlinecite{Gorkov} and Abrahams et al.\ in Ref.\ \onlinecite{Abrahams}, who realized that the leading quantum correction to the classical diffusive conductivity is given by the sum of maximally crossed diagrams, which diverges in one and two dimensions. It was later understood by Bergmann,\cite{Bergmann} Khmel'nitskii and Larkin,\cite{Larkin} and Chakravarty and Smith \cite{Chakravarty} that the weak localization correction is related to the phenomenon of coherent backscattering. Sensitivity of localization to the spin-orbit coupling and magnetic field and the emerging magnetoresistance was first discussed by Hikami et al.\ in Ref.\ \onlinecite{Hikami}. These seminal works laid the foundadtion of a whole new branch of mesoscopic physics and have stimulated extensive experimental and theoretical studies of coherent transport properties of metals and semiconductors (for a review, see Ref.\ \onlinecite{Altshuler12}).

Interference effects are particularly pronounced in low-dimensional samples since electron trajectories have much stronger tendency to self-intersection. When the transverse size of the sample $W$ is particularly small (comparable or less than the electron mean free path $l$), transport properties are sensitive to the boundary conditions and, in particular, to boundary scattering. Magnetoresistance in parallel magnetic field is also affected by the properties of the sample boundary provided the magnetic length $l_B = \sqrt{\hbar c/e B}$ is larger or comparable to $W$. This situation is often realized in semiconducting nanowires, exhibiting both negative \cite{Kurdak, Katine, Lettau, Koester, Niimi, Reulet} and positive \cite{Kurdak, Schaepers, Hansen, Kallaher, Kallaher2} magnetoresistance and multiwall carbon nanotubes.\cite{Schoenenberger, Liu, Stojetz, Strunk, Kang} Recently, similar transport regime was also encountered in the experiments on topological insulators, that are materials with an extremely strong spin-orbit coupling, in thin films of BiSe (see Refs.\ \onlinecite{Chen, Checkelsky, Lin}), BiTe (see Ref.\ \onlinecite{Chiu}), and in the HgTe nanowires.\cite{Muehlbauer}

Parallel field magnetoresistance in a relatively thick ($l_B \gtrsim W \gg l$) quasi-2D film was first considered theoretically by Altshuler and Aronov.\cite{Altshuler81} In such samples, transverse electron propagation is of the diffusive character and magnetoresistance is insensitive to boundary scattering. Later, Dugaev and Khmel'nitskii \cite{Dugaev} studied the opposite (ballistic) limit $W \ll l$ with diffuse boundary and identified two qualitatively different limits of weak ($l_B \gg \sqrt{W l}$) and strong ($l_B \ll \sqrt{W l}$) magnetic field. Beenakker and van Houten \cite{Beenakker} further extended the theory including both the wire (quasi-1D) and the film (quasi-2D) geometries with either mirror or diffuse boundaries. They obtained analytic expressions for the ballistic and diffusive limits and some numerical results for the crossover between them.  Samples of the nanotube geometry were considered in the work by Takane \cite{Takane} where analytic expressions for magnetoresistance were derived for the three asymptotic regimes (diffusive and ballistic in either weak or strong magnetic field). 

Existing theoretical results are however insufficient to describe the variety of experimental situations, where it is quite common to have the width of the sample comparable to the electron mean free path, so that the sample is in the crossover between diffusive and ballistic limits. In particular, this is the case in experiments on semiconductor nanowires,\cite{Katine, Reulet, Hansen, Schaepers, Kallaher} carbon nanotubes,\cite{Schoenenberger, Liu, Stojetz, Strunk, Kang} and topological insulators.\cite{Chiu, Lin, Muehlbauer} It was already noted in Refs.\ \onlinecite{Hansen, Schoenenberger} that fitting experimental data to the Beenakker-van Houten theory \cite{Beenakker, Takane} in the pure ballistic or diffusive regime does not produce a good agreement due to the fact that $W \sim l$. Crossover from weak ($l_B \gg \sqrt{Wl}$) to strong ($l_B \ll \sqrt{Wl}$) magnetic field in ballistic samples is also easily accessible in the experiments. The lack of analytic results in the intermediate field range is usually circumvented by interpolation between the two asymptotic formulas.\cite{Kurdak, Katine, Koester} Pure mirror or diffuse boundary conditions are also not always an adequate description for experiments. The reason is that in many cases,\cite{Katine, Niimi} the magnetotransport data can be fitted relatively well within the mirror boundary model while the diffusion coefficient still depends on the sample width suggesting a significant randomness of the boundary scattering. 

The purpose of this paper is to overcome the limitations of the previous studies by deriving the most general analytic expressions for magnetoresistance suitable for all the experimentally relevant regimes. First, our results fully cover the crossover between diffusive and ballistic limits allowing for an arbitrary ratio between $W$ and $l$. Second, we consider mixed boundary scattering \cite{Fuchs} with the probabilities $\lambda$ and $1 - \lambda$ of specular and diffuse reflection, respectively. This allows us to fill the gap between the two idealized models and provides an additional parameter important for explaining experimental data. We also discuss periodic boundary conditions and compute magnetoresistance of the nanotube in perpendicular magnetic field in the whole range of $W/l$ ratio. Thus our analytic results cover all the limiting cases of both diffusive and ballistic samples in weak or strong magnetic fields including possible crossovers between them and allowing for general boundary scattering. In particular, we demonstrate that in a certain parameter range, the electron mobility can be limited by the boundary scattering while the magnetoresistance is only weakly sensitive to the edge roughness, thereby suggesting a resolution to the discrepancy observed in Refs.\ \onlinecite{Katine, Niimi}.

There are two equivalent techniques to compute weak localization corrections and magnetoresistance. The quantum correction to conductance can be represented as a sum of maximally crossed diagrams \cite{Gorkov, Abrahams}. Such a diagrammatic calculation accurately accounts for all the microscopic details of the system but is quite tedious in the presence of a complicated boundary. For this reason, an alternative, and more intuitive, formalism of Boltzmann kinetic equation \cite{Dugaev, Beenakker} is usually applied when boundary conditions are important. This approach is based on the relation between the quantum correction to conductivity and the classical probability of return. Finally, a third alternative method used by Beenakker and van Houten \cite{Beenakker} is based on the analysis of classical trajectories. This method was first proposed in an earlier work by de Gennes and Tinkham.\cite{DeGennes}

The paper is organized as follows. In Sec.\ \ref{Sec:formalism}, we describe the general model of a nanowire and explain the details of the kinetic equation technique with different boundary conditions. Classical diffusion and its dependence on boundary conditions is discussed in Sec.\ \ref{Sec:diffusion}. The next two sections\ \ref{Sec:weak} and \ref{Sec:strong} are devoted to the solution of the problem in the limits of weak ($l_B \gg \sqrt{Wl}$) and strong ($l_B \ll \sqrt{Wl}$) magnetic field, respectively. General results for a sample of quasi-1D geometry are presented in Sec.\ \ref{Sec:q1d}. In Sec.\ \ref{Sec:q2d}, we discuss the magnetoconductivity of thin metallic films. The concluding Sec.\ \ref{Sec:discussion} summarizes the main findings of the paper, see Table \ref{tsumm}. Explicit computation of the Cooperon is presented in Appendix \ref{App:Cooperon}. The diagrammatic derivation of the quantum correction is given in Appendix \ref{App:diag}.

\section{Kinetic equation formalism}
\label{Sec:formalism}

\subsection{Model and parameters}

We consider a disordered 2D electron system that is patterned in a shape of a long quasi-1D channel along the $x$ direction with width $W$ along the $y$ direction. Electron motion in the $y$ direction is restricted by the boundaries at $y = \pm W/2$. The prototypes for our model are semiconducting heterostructures hosting a conducting quasi-1D channel in a quantum well.\cite{Kurdak, Katine, Lettau, Koester, Niimi, Reulet, Schaepers, Hansen, Kallaher, Kallaher2, Muehlbauer} We assume metallic regime of electron transport with Fermi velocity $v_0$ and Fermi wavelength much shorter than all other length scales. The disorder is described within the Gaussian white noise model with the mean free time $\tau$ and the mean free path $l = v_0 \tau$. Additional contribution to electron scattering is provided by the boundaries of the channel. An electron incident at the edge of the channel is either reflected in a random direction with probability $1 - \lambda$ or experiences mirror reflection with probability $\lambda$. External magnetic field $B$ is applied in the direction perpendicular to the 2D electron gas. 

There are in total three relevant length scales in the problem: the channel width $W$, electron mean free path $l$, and magnetic length $l_B = \sqrt{\hbar c/ e B}$. The latter is the characteristic length at which the electron wave function picks a significant phase due to the magnetic field. In addition, the temperature-dependent electron dephasing length $l_\phi$, which arises, e.g., due to Coulomb interaction, limits the range of coherent transport and suppresses the weak localization effects. Throughout the paper we assume the temperature is sufficiently low and hence $l_\phi$ is longer than the other three length scales. In samples with strong spin-orbit coupling, an additional coherent spin relaxation length $l_\text{SO}$ emerges. We consider only the two extreme cases of either negligible ($l_\text{SO} > l_\phi$, weak localization) or very strong ($l_\text{SO} < l$, weak antilocalization) spin-orbit coupling.

The ratios of the main three length scales determine the regime of electron transport and the magnetoresistance. We assume $l_B \gg W$, that is the condition of quasi-1D magnetotransport. This condition is always fulfilled at sufficiently low magnetic fields. The ratio $W/l$ distinguishes between ballistic ($W \ll l$) and diffusive ($W \gg l$) transport regimes in the transverse direction. We will consider arbitrary values of $W/l$ and obtain a general expression for megnetoconductivity, that reduces to earlier results in the diffusive \cite{Altshuler81} and ballistic \cite{Dugaev, Beenakker} limits. For ballistic samples, $W \ll l$, the cases of relatively weak and strong magnetic field are discriminated by the ratio $l_B/\sqrt{W l}$. In a weak field, $l_B \gg \sqrt{W l}$, quantum interference is dominated by the returning trajectories with a large number of impurity scatterings while for $W \ll l_B \ll \sqrt{W l}$ only loops with a few impurities are important.

\subsection{Kinetic equation}

Electron dynamics in a disordered system is governed by the diffusion equation at large scales. This results in the classical Drude value of the conductivity $\sigma_0 = e^2 \nu D$, where $\nu$ is the electron density of states at the Fermi level and $D$ is the diffusion coefficient. Quantum correction to this classical value arises due to interference between electron trajectories and is given by the Cooperon at coincident points:\cite{Altshuler81, Dugaev}
\begin{equation}
 \Delta \sigma
  = \eta \frac{\sigma_0}{\pi \nu \hbar} \langle \mathcal{C}(\mathbf{r}, \mathbf{r}) \rangle.
 \label{Ds}
\end{equation}
Here the factor $\eta$ equals $-2$ for systems without spin-orbit coupling (weak localization) and $+1$ when the spin-orbit coupling is strong (weak antilocalization). The fact that the prefactor of Eq.\ (\ref{Ds}) is always given by the Drude conductivity regardless of the system geometry or boundary scattering is justified by the semiclassical treatment of Refs.\ \onlinecite{Chakravarty, Beenakker}. In Appendix \ref{App:diag}, we show that this is the case by explicitly performing the diagrammatic analysis.

The Cooperon $\mathcal{C}$ is the propagator of the Boltzmann kinetic equation. In a symbolic form it can be written as
\begin{equation}
\label{BE}
 \left[
   \frac{1}{\tau_\phi} + \mathbf{v} \left( \nabla - \frac{2i e}{\hbar c} \mathbf{A} \right) + \frac{1 - | 0 \rangle \langle 0 |}{\tau} 
 \right] \mathcal{C}
  = \mathbbm{1}.
\end{equation}
The propagator $\mathcal{C}$ is the operator inverse of the expression in brackets. This operator acts in the phase space of position and momentum. Since in a metallic state, relevant momenta are restricted to the Fermi surface, we use the velocity vector $\mathbf{v}$ (with the absolute value of Fermi velocity $v_0$) instead of momentum to label the position on the Fermi surface. Hence, when written explicitly, $\mathcal{C}$ depends on two points in real space, $\mathbf{r}$ and $\mathbf{r}'$, and two points on the Fermi surface $\mathbf{v}$ and $\mathbf{v}'$. The operator $| 0 \rangle \langle 0 |$ averages over directions of $\mathbf{v}$ or, equivalently, projects on to the constant function $| 0 \rangle$ over the Fermi surface. In the case of a circular 2D Fermi surface, this is just an integral operator $| 0 \rangle \langle 0 | = \int d\phi/2\pi$ and the right-hand side of Eq.\ (\ref{BE}) is $\mathbbm{1} = 2\pi \delta(\phi - \phi') \delta(\mathbf{r} - \mathbf{r}')$ with $\phi$ being the direction of velocity. We will compute the quantum correction to conductivity, Eq.\ (\ref{Ds}), at zero frequency, hence we have omitted the frequency term in Eq.\ (\ref{BE}). At the same time, a finite dephasing rate $1/\tau_\phi$ is included to account for the coherence breaking effects.

The quantum correction (\ref{Ds}) is given by the Cooperon at coincident points averaged over both directions $\mathbf{v}$ and $\mathbf{v}'$. This is exactly the probability to return to the same point in space (regardless of the velocity direction) during semiclassical motion described by the kinetic equation. An external magnetic field enters Eq.\ (\ref{BE}) via the vector potential $\mathbf{A}$. This corresponds to a relative phase acquired by two electrons following the same trajectory in opposite directions. Hence the name Cooperon and the charge $2e$ in the vector potential coupling. It is this phase that destroys coherence and leads to magnetoresistance.

\subsection{Boundary conditions}
\label{Sec:boundary}

Let us now discuss possible boundary conditions to the Boltzmann equation (\ref{BE}) at the edges of the conducting channel $y = \pm W/2$. 

\paragraph*{Mirror boundary.} The simplest approximation is the flat mirror boundary. In this case, an electron approaching the edge of the sample with the velocity directed at an angle $\phi$ (relative to the longitudinal $x$ direction) is reflected at the angle $-\phi$. Hence the Cooperon propagator obeys
\begin{equation}
 \mathcal{C}(y = \pm W/2, \phi)
  = \mathcal{C}(y = \pm W/2, -\phi).
 \label{mirror}
\end{equation}
For brevity, we show explicitly only the relevant arguments of the function $\mathcal{C}$: the $y$ component of the position $\mathbf{r}$ and the direction of $\mathbf{v}$. The condition (\ref{mirror}) implies that the densities of electrons with velocity directions $\pm \phi$ are equal at $y = \pm W/2$.

\paragraph*{Diffuse boundary.} In the opposite case of a rough disordered edge, the boundary conditions acquire the diffuse form. Irrespective of the incidence angle, the electron is reflected in all directions with equal probability. Such boundary conditions have the form
\begin{equation}
 \mathcal{C}(y = \pm W/2, \mp \phi)
  = \int\limits_0^\pi d\phi_1 \frac{\sin\phi_1}{2} \mathcal{C}(y = \pm W/2, \pm \phi_1).
 \label{diffuse}
\end{equation}
Here the angle $\phi$ is restricted to the range $[0, \pi]$. The boundary conditions imply that the electron flow reflected from the edge at an angle $\mp \phi$ is proportional to the total incident flow integrated over all incident angles $\pm \phi_1$. The $\sin \phi_1$ factor in the integrand accounts for the angular dependence of the incidence rate.

\paragraph*{Mixed boundary.} The boundaries of a real material are somewhere between the two idealized limits of mirror and diffuse edge. The most general boundary condition, for which an analytic solution is still feasible, is the mixture of the two cases discussed above.\cite{Fuchs} In this general case, an incident electron has probabilities $\lambda$ and $1 - \lambda$ of mirror and diffuse reflection, respectively. The boundary condition is the corresponding linear combination of Eqs.\ (\ref{mirror}) and (\ref{diffuse}):
\begin{multline}
 \mathcal{C}(y = \pm W/2, \mp \phi)
  = \lambda\, \mathcal{C}(y = \pm W/2, \pm \phi) \\
    + \frac{1 - \lambda}{2} \int\limits_0^\pi d\phi_1\, \sin\phi_1\, \mathcal{C}(y = \pm W/2, \pm \phi_1).
 \label{mixed}
\end{multline}

\paragraph*{Periodic boundary conditions.} If the quasi-1D system has the form of a cylinder, that is the case for carbon nanotubes, \cite{Schoenenberger, Liu, Stojetz, Strunk, Kang} boundary conditions are periodic
\begin{equation}
 \mathcal{C}(y = W/2, \phi)
  = \mathcal{C}(y = -W/2, \phi).
\end{equation}
Here we assume that $y$ measures the distance along the nanotube circumference and the width $W = 2\pi R$ is determined by the tube radius $R$.

Since we aim to solve Eq.\ (\ref{BE}) for a quasi-1D system, the Landau gauge for the vector potential $\mathbf{A}$ is most convenient. In this gauge, only the longitudinal component $A_x$ is nonzero. In a uniform perpendicular magnetic field $B$, the vector potential is $A_x = B y$ for a flat quasi-1D conducting channel and $A_x = B R \cos(y/R)$ in the case of a cylinder.

\subsection{General solution}
\label{Sec:general}

The kinetic equation for the Cooperon can be solved in two steps.\cite{Dugaev} First, we disregard the projection term $|0 \rangle \langle 0 |$ in the left-hand side of Eq.\ (\ref{BE}) and introduce the operator $\mathcal{C}_0$ that solves the reduced equation:
\begin{equation}
 \mathcal{C}_0
  = \left[
      \frac{1}{\tau} + \frac{1}{\tau_\phi} + \mathbf{v} \left( \nabla - \frac{2i e}{\hbar c} \mathbf{A} \right)
    \right] ^{-1}.
 \label{C0}
\end{equation}
This operator acts in the same phase space of position and momentum as the full Cooperon $\mathcal{C}$ and obeys the same boundary conditions at $y = \pm W/2$. The right-hand side of Eq.\ (\ref{C0}) is diagonal in momentum (velocity) subspace. However, boundary conditions violate the conservation of momentum (except for the periodic conditions in case of a nanotube) and, hence, $\mathcal{C}_0$ is actually not diagonal in velocity. At the second step, solution to the full equation (\ref{BE}) can be written in terms of $\mathcal{C}_0$ as
\begin{multline}
 \mathcal{C}
  = \left[
      \mathcal{C}_0^{-1} - \tau^{-1} | 0 \rangle \langle 0 |
    \right]^{-1} \\
  = \mathcal{C}_0 + \mathcal{C}_0 | 0 \rangle \bigl( \tau - \langle 0 | \mathcal{C}_0 | 0 \rangle \bigr)^{-1} \langle 0 | \mathcal{C}_0.
 \label{GS}
\end{multline}
Computation of $\mathcal{C}$ amounts to inverting the operator $\tau - \langle 0 | \mathcal{C}_0 | 0 \rangle$ in real space only. We note that the expression (\ref{GS}) has the form of a geometric series corresponding to the ladder diagrams with $\mathcal{C}_0$ being a single ladder rung. In Appendix \ref{App:diag}, we reproduce all the results within the microscopic calculation of such ladder diagrams.

For a quasi-1D system, both Eqs.\ (\ref{BE}) and (\ref{C0}) are translationally invariant along $x$. Switching over to momentum representation $\mathcal{C}_0(x,x') = \exp[i q (x - x')] \mathcal{C}_0(q)$, we reduce Eq.\ (\ref{C0}) to a differential equation in $y$ only. Explicitly,
\begin{equation}
 \left\{
   \frac{1}{\tau} + \frac{1}{\tau_\phi} + v_0 \sin\phi \frac{\partial}{\partial y} + i v_0 \cos\phi \left[ q - \frac{2 e}{\hbar c} A_x(y) \right]
 \right\} \mathcal{C}_0
  = \mathbbm{1}.
 \label{C0q}
\end{equation}
Quantum correction to quasi-1D conductivity is given by Eq.\ (\ref{Ds}) integrated over $y$. With the help of Eq.\ (\ref{GS}), we can express the result directly in terms of the angular average $\langle 0 | \mathcal{C}_0 | 0 \rangle$:
\begin{equation}
\label{Ds1D}
 \Delta \sigma_\text{1D}
  = \eta \frac{e^2 D \tau}{\pi \hbar} \int\frac{dq}{2\pi} \int\limits_{-W/2}^{W/2} \frac{dy}{W}
      \bigl[ 1 - \tau^{-1}\langle 0 | \mathcal{C}_0 | 0 \rangle \bigr]^{-1}_{y,y}.
\end{equation}
Here, we have omitted a constant term in the integrand since it does not contribute to the magnetic field dependence of the quantum correction.

The main contribution to the magnetoconductance comes from the domain $q \ll 1/l$ where the integrand in Eq.\ (\ref{Ds1D}) is particularly large. To demonstrate this, let us for a moment neglect the dephasing rate and magnetic field, and assume $q = 0$. Under these assumptions, the operator $\mathcal{C}_0$ has an eigenvalue $\tau$. The corresponding eigenfunction is constant in both $y$ and $\phi$. The angular averaged operator $\langle 0 | \mathcal{C}_0 | 0 \rangle$ possesses the same eigenvalue, hence, the integrand in Eq.\ (\ref{Ds1D}) diverges. Such divergence is known as the diffusion pole; it is related to the conservation of the total number of electrons. The pole appears in the Cooper channel due to the time-reversal symmetry in the absence of magnetic field.

The smallest nonzero eigenvalue of the denominator in Eq.\ (\ref{Ds1D}) corresponds to an eigenfunction varying in $y$ on the scale $W$ and can be estimated as $l/W$. The contribution of a finite magnetic field to the lowest (zero) eigenvalue is of the order $e B l W/\hbar c \sim l W/l_B^2$, where $l_B = \sqrt{\hbar c/e B}$ is the magnetic length. Hence magnetic field produces a negligible effect on all nonzero eigenvalues in the quasi-1D limit $l_B \gg W$. For the calculation of magnetoresistance it is thus sufficient to keep only the lowest eigenstate of $\langle 0 | \mathcal{C}_0 | 0 \rangle$ and further simplify Eq.\ (\ref{Ds1D}) by projecting on to constant function in $y$.
\begin{gather}
 \Delta \sigma_\text{1D}
  = \eta \frac{e^2 D \tau}{\pi h} \int \frac{dq}{1 - \tau^{-1} \langle\!\langle \mathcal{C}_0(q) \rangle\!\rangle}, \label{Dsigma} \\
 \langle\!\langle \mathcal{C}_0(q) \rangle\!\rangle
  = \int\limits_{-W/2}^{W/2} \frac{dy\, dy'}{W^2} \int \frac{d\phi\, d\phi'}{(2\pi)^2}\, \mathcal{C}_0(q, y, \phi, y', \phi'). \label{avC}
\end{gather}

We are now in the position to expand the averaged operator $\langle\!\langle \mathcal{C}_0(q) \rangle\!\rangle$ in small momentum $q$ and dephasing rate $1/\tau_\phi$. This expansion can be written in terms of the reduced operator
\begin{equation}
 \mathcal{R}
  = \left[
      1 + l \sin\phi \frac{\partial}{\partial y}
    \right] ^{-1}
 \label{R}
\end{equation}
that obeys the same boundary conditions as the operators $\mathcal{C}$ and $\mathcal{C}_0$; in Appendix \ref{App:Cooperon}, we compute $\mathcal{R}$ for different boundary conditions explicitly. Dependence of the Cooperon on magnetic field can be singled out with the help of the following transformation:
\begin{gather}
 \mathcal{C}_0
  = e^{i \alpha \cot\phi} \left(
      \frac{\mathcal{R}^{-1}}{\tau} + \frac{1}{\tau_\phi} + i v_0 q \cos\phi
    \right)^{-1} e^{-i \alpha \cot\phi}, \label{C0R} \\
 \alpha(y)
  = \frac{2 e}{\hbar c} \int\limits_{-W/2}^y A_x(y')\, dy'. \label{alpha}
\end{gather}
We assume that a constant in the definition of the vector potential is fixed such that $\alpha(y = \pm W/2) = 0$ and hence magnetic field does not alter the boundary conditions.

In the absence of dephasing and at $q = 0$, the denominator of Eq.\ (\ref{Dsigma}) determines the magnetic scattering rate 
\begin{equation}
 \frac{1}{\tau_B}
  = \frac{1 - \langle\!\langle e^{i\alpha\cot\phi} \mathcal{R} e^{-i\alpha\cot\phi} \rangle\!\rangle}{\tau}.
 \label{tauB}
\end{equation}
Similar to $\mathcal{C}_0$, the operator $\mathcal{R}$ has an eigenvalue $1$ corresponding to the constant eigenfunction. Hence the magnetic scattering rate vanishes in the absence of magnetic field and the diffusion pole emerges. With the help of Eq.\ (\ref{C0R}), we expand $\langle\!\langle \mathcal{C}_0(q) \rangle\!\rangle$ to the second order in $q$ and to the first order in $1/\tau_\phi$. Magnetic field is neglected ($\alpha$ is set to zero) in all but the lowest term of this expansion. Higher terms contain averages of the form $\langle\!\langle \mathcal{R} \ldots \mathcal{R} \rangle\!\rangle$. Since the constant function is invariant under the action of $\mathcal{R}$, we can omit the first and the last operators in these averages. This yields the following expression for the denominator of Eq.\ (\ref{Dsigma}):
\begin{equation}
 1 - \tau^{-1} \langle\!\langle \mathcal{C}_0(q) \rangle\!\rangle
  = \tau (D q^2 + \tau_\phi^{-1} + \tau_B^{-1}),
 \label{denominator}
\end{equation}
where the diffusion coefficient is given by
\begin{equation}
 D
  = v_0 l \langle\!\langle \cos\phi\; \mathcal{R}\; \cos\phi \rangle\!\rangle.
 \label{D}
\end{equation}

With the expansion (\ref{denominator}), we carry out momentum integration in Eq.\ (\ref{Dsigma}) and obtain magnetoconductivity in the form
\begin{equation}
\label{MR}
 \Delta \sigma_\text{1D}(B)
  = \eta \frac{e^2}{h} \left[ \sqrt\frac{D}{\tau_B^{-1} + \tau_\phi^{-1}} - \sqrt{D \tau_{\phi}} \right],
\end{equation}
where we have subtracted the zero field contribution such that the magnetoconductance is measured relative to the value at zero field. To summarize, the general result for magnetoconductance is expressed in terms of the diffusion coefficient $D$ and the magnetic scattering rate $1/\tau_B$ defined by Eqs.\ (\ref{D}) and (\ref{tauB}). Boundary properties are encoded in the operator $\mathcal{R}$ and influence the values of both $D$ and $\tau_B$ while magnetic field enters only the latter quantity via the phase $\alpha$ defined by Eq.\ (\ref{alpha}).

\subsection{Weak field limit}
\label{Sec:weakfield}

The calculation of magnetoconductance can be simplified in the limit of relatively weak magnetic field. The term with vector potential in Eq.\ (\ref{C0}) is small compared to $1/\tau$ provided $l_B \gg \sqrt{W l}$. In this case, perturbative expansion in magnetic field is valid. Expanding the phase factors in Eq.\ (\ref{tauB}) in powers of $\alpha$ and using the fact that a constant is the eigenfunction of $\mathcal{R}$ with the eigenvalue $1$, we express the magnetic scattering rate as
\begin{equation}
\label{tBw}
 \frac{1}{\tau_B}
  = \frac{1}{\tau}\langle\!\langle \alpha \cot\phi (\mathbbm{1} - \mathcal{R}) \alpha \cot\phi \rangle\!\rangle. 
\end{equation}

This result can be simplified even further for a thick sample, $W \gg l$. The transverse motion of electrons is of diffusive type in this case and the boundary conditions are unimportant. Technically, this corresponds to the fact that the operator $\mathcal{R}$ is close to $1$ and can be expanded in powers of gradient. This yields the magnetic scattering rate in the form \cite{Altshuler81}
\begin{equation}
 \frac{1}{\tau_B}
  = v_0 l \left<\!\!\left< \Bigl(\cos\phi\; \frac{\partial\alpha}{\partial y} \Bigr)^2 \right>\!\!\right>
  = \frac{2e^2 l^2}{\hbar^2 c^2 \tau} \langle\!\langle A_x^2 \rangle\!\rangle,
\end{equation}
while the diffusion coefficient, Eq.\ (\ref{D}), takes its standard value $D = v_0 l/2$. For the flat and cylinder samples, we have
\begin{equation}
 \label{weakdiffuse}
 \frac{1}{\tau_B}
  = \frac{l^2 W^2}{\tau l_B^4} \begin{cases}
      1/6 , & \text{flat}, \\
      1/4 \pi^2, &\text{cylinder}.
    \end{cases}
\end{equation}

Thus, there are in total three limits of qualitatively different magnetoresistance: diffusive sample ($l_B \gg W \gg l$), ballistic sample in a weak magnetic field ($l_B \gg \sqrt{W l} \gg W$), and ballistic sample in a relatively strong magnetic field ($\sqrt{W l} \gg l_B \gg W$). The results in all these three cases are summarized in the end of the paper in Table \ref{tsumm}.

\section{Magnetoconductivity of quasi-1D samples}

In this Section we compute the diffusion coefficient $D$, Eq.\ (\ref{D}), and the magnetic scattering rate $\tau_B^{-1}$, Eq.\ (\ref{tauB}), in a quasi-1D sample with the particular boundary conditions discussed in Sec.\ \ref{Sec:boundary}. These two parameters determine the magnetoconductivity via Eq.\ (\ref{MR}). In the subsequent calculations, we use the explicit expression for the reduced Cooperon $\mathcal{R}$, Eqs.\ (\ref{Rf}) and (\ref{Rc}), computed in Appendix \ref{App:Cooperon}.

\subsection{Diffusion coefficient}
\label{Sec:diffusion}

\begin{figure}
\centering
\includegraphics[width = 0.95\columnwidth]{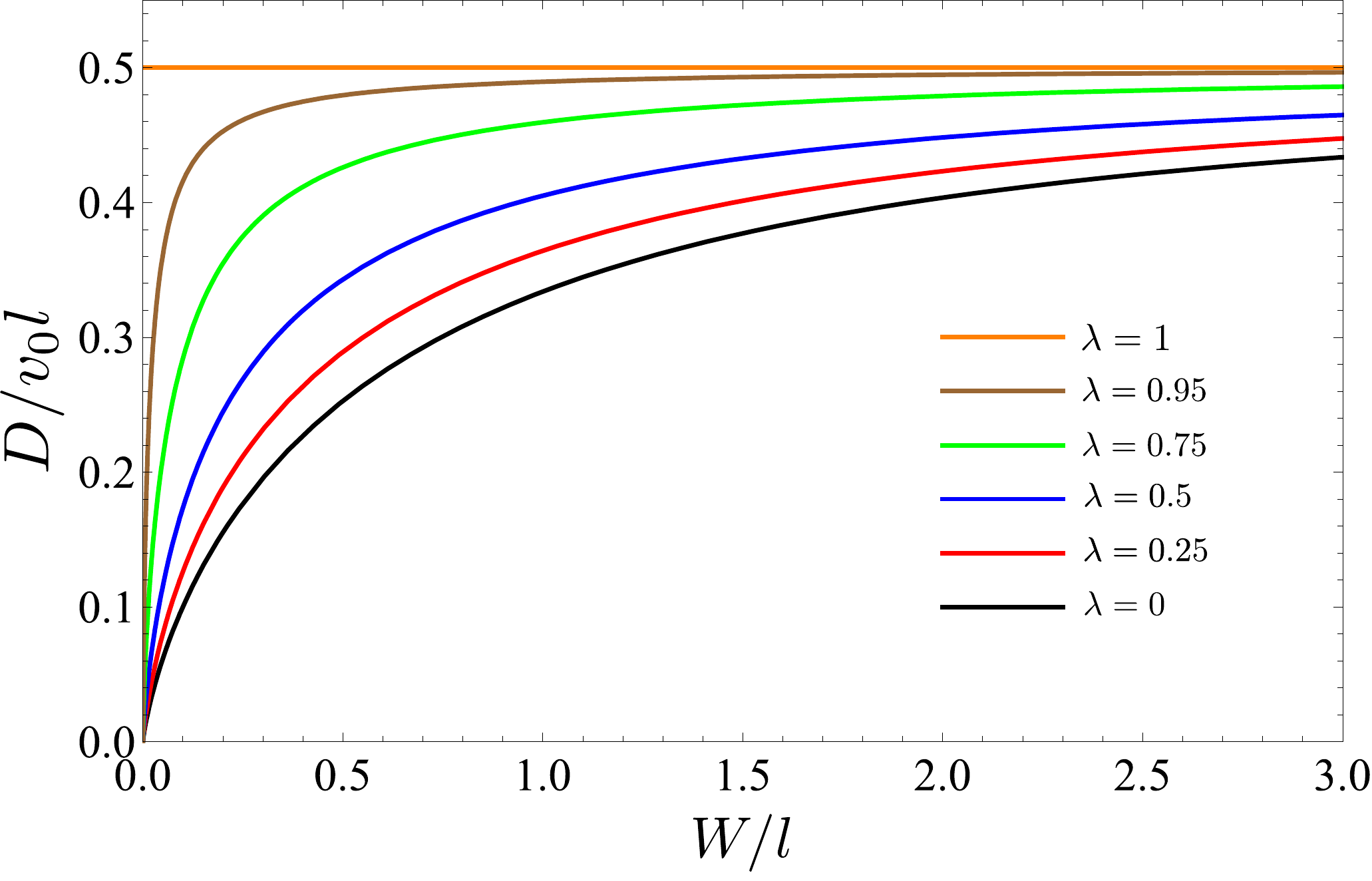}
\caption{Diffusion coefficient in units of $v_0 l$ as a function of $W/l$ for different values of $\lambda$ computed using Eq.\ (\protect\ref{Dm}).}
\label{Dmf}
\end{figure}

The diffusion coefficient is given by Eq.\ (\ref{D}) where double angular brackets imply integration with respect to transverse coordinates $y$ and velocity directions $\phi$ on both sides of the reduced Cooperon propagator $\mathcal{R}$. In the cases of mirror or periodic boundaries, the diffusion coefficient is not influenced by the boundary scattering and is the same as in an infinite two dimensional system $D = v_0 l/2$. In the case of a planar quasi-1D channel with the general mixed boundary conditions (\ref{mixed}), we use the general expression (\ref{Rf}). Spatial dependence of the reduced Cooperon is given by the exponential factor as in Eq.\ (\ref{Ry}). This allows us to integrate over $y$ and $y'$ first. Subsequent angular integration cancels the first term in Eq.\ (\ref{Rf}) since it is invariant under $\phi \mapsto \pi - \phi$ and hence vanishes when integrated with the weight $\cos \phi$. The diffusion coefficient is thus determined by the last two terms of Eq.\ (\ref{Rf}). Due to the factors $\delta(\phi \pm \phi')$, we obtain the expression for the diffusion coefficient in the form of a single angular integral
\begin{multline}
 \label{Dm}
 D
  = \frac{v_0 l}{2} \left[ \rule{0pt}{7mm} 1 - (1 - \lambda) \frac{2 l}{\pi W} \right. \\ \left.
      \times \int_0^{\pi} d\phi\, \sin \phi\, \cos^2 \phi\,
      \frac{1 - e^{-\tfrac{W}{l \sin \phi}}}{1 - \lambda e^{-\tfrac{W}{l \sin \phi}}}
    \right].
\end{multline}
This expression reduces to $v_0 l/2$ in the case of mirror boundary, $\lambda = 1$, and reproduces the result of Ref.\ \onlinecite{Beenakker} in the opposite case of purely diffuse boundaries, $\lambda = 0$. The dependence of the diffusion coefficient on $W/l$ for a few values of $\lambda$ is shown in Fig.\ \ref{Dmf}. We notice that for any $\lambda < 1$, the diffusion coefficient decreases for sufficiently small $W$. This means that deep in the ballistic regime $W \ll l$, the pure mirror boundary is not a good approximation. The reason for this is that in a very narrow sample the particle undergoes multiple boundary scatterings before it is scattered by a bulk impurity. So even a small probability of diffuse scattering at the boundary is effectively enhanced.

Let us analyze the limiting behavior of the diffusion coefficient. For a wide sample, $W \gg l$, we can neglect exponential terms in the integrand of Eq.\ (\ref{Dm}) and obtain
\begin{equation}
 \label{Ddif}
 D
  = \frac{v_0 l}{2} \left[
      1 - \frac{4 (1 - \lambda) l}{3 \pi W}
    \right].
\end{equation}
The second term in the brackets provides a small correction to the diffusion coefficient due to scattering at the boundary. All higher order corrections are exponentially small in $W/l$. In the opposite limit of a narrow sample, $W \ll l$, two qualitatively different transport regimes arise depending on the parameter
\begin{equation}
 \label{b}
 b
  = \frac{2(1 - \lambda)l}{(1 + \lambda)W}.
\end{equation}
When $b$ is large, the dominant scattering mechanism is due to diffuse scattering at the boundary while for $b \ll 1$ the mobility is limited by the bulk scattering. Exponential terms in Eq.\ (\ref{Dm}) can be expanded provided $\phi$ is not very close to $0$ or $\pi$ yielding
\begin{multline}
 \label{Dphi}
 D
  = \frac{v_0 l}{\pi} \int_0^{\pi} \frac{d\phi\, \cos^2 \phi}{1 + b \sin \phi} \\
  = \frac{v_0 l}{b^2} \begin{dcases}
      1 - \frac{2}{\pi} \bigl( b - \sqrt{b^2 - 1} \mathop{\mathrm{arccosh}} b \bigr), & b > 1, \\
      1 - \frac{2}{\pi} \bigl( b + \sqrt{1 - b^2} \arccos b \bigr), & b < 1.
    \end{dcases}
\end{multline}
In the asymptotic regimes of predominantly boundary ($b \gg 1$) or bulk ($b \ll 1$) scattering this crossover expression reduces to
\begin{equation}
 \label{Dcross}
 D
  = v_0 l \begin{dcases}
      \frac{(1 + \lambda)W}{(1 - \lambda) \pi l} \ln \frac{(1 - \lambda)l}{(1 + \lambda)W}, & W/l \ll 1 - \lambda, \\
      \frac{1}{2} - \frac{2 (1 - \lambda)l}{3\pi W}, & 1 - \lambda \ll W/l \ll 1.
    \end{dcases}
\end{equation}
Remarkably, the result (\ref{Ddif}) for a wide sample coincides with the second case of the above expression including the $1 - \lambda$ correction term. This suggests that we can consider the whole range $W/l \gg 1 - \lambda$ as the single limit of wide sample at least with respect to the diffusion coefficient. In this limit, the mobility of electrons is limited by the bulk scattering.

For a narrow sample with almost mirror boundaries, $W/l \ll 1 - \lambda \ll 1$, the integral in Eq.\ (\ref{Dphi}) is mainly determined by small angles $\phi \lesssim W/[l (1 - \lambda)]$. Such shallow trajectories typically scatter $1/(1 - \lambda) \gg 1$ times on the length $l$ between two successive bulk scatterings. This explains the high sensitivity of the diffusion constant to the roughness of sample edges in the ballistic limit $W \ll l$. The crossover from specular to effectively diffuse boundary scattering occurs for a parametrically small roughness $1 - \lambda \sim W/l$.

\subsection{Magnetic scattering rate in a weak magnetic field $l_B \gg \sqrt{W l}, W$}
\label{Sec:weak}

In the limit of weak magnetic field $l_B \gg \sqrt{W l}$, the computation of the magnetic scattering rate is simplified as explained in Sec.\ \ref{Sec:weakfield}. The reason for this simplification is that the phase acquired by an electron due to the magnetic field between two impurity scatterings is small. The magnetic scattering rate is obtained by substituting the reduced Cooperon from Eqs.\ (\ref{Rf}) or (\ref{Rc}) into Eq.\ (\ref{tBw}). In both cases of cylinder and flat sample, the result can be represented in the form
\begin{equation}
 \label{tby}
 \frac{1}{\tau_B}
  = \frac{W^2 l^2}{\tau l_B^4}\, g(W/l)
\end{equation}
where the function $g(x)$ encodes all the information about the geometry of the sample and the boundary conditions.

\subsubsection{Cylinder sample}

\begin{figure}
\centering
\includegraphics[width = 0.9\columnwidth]{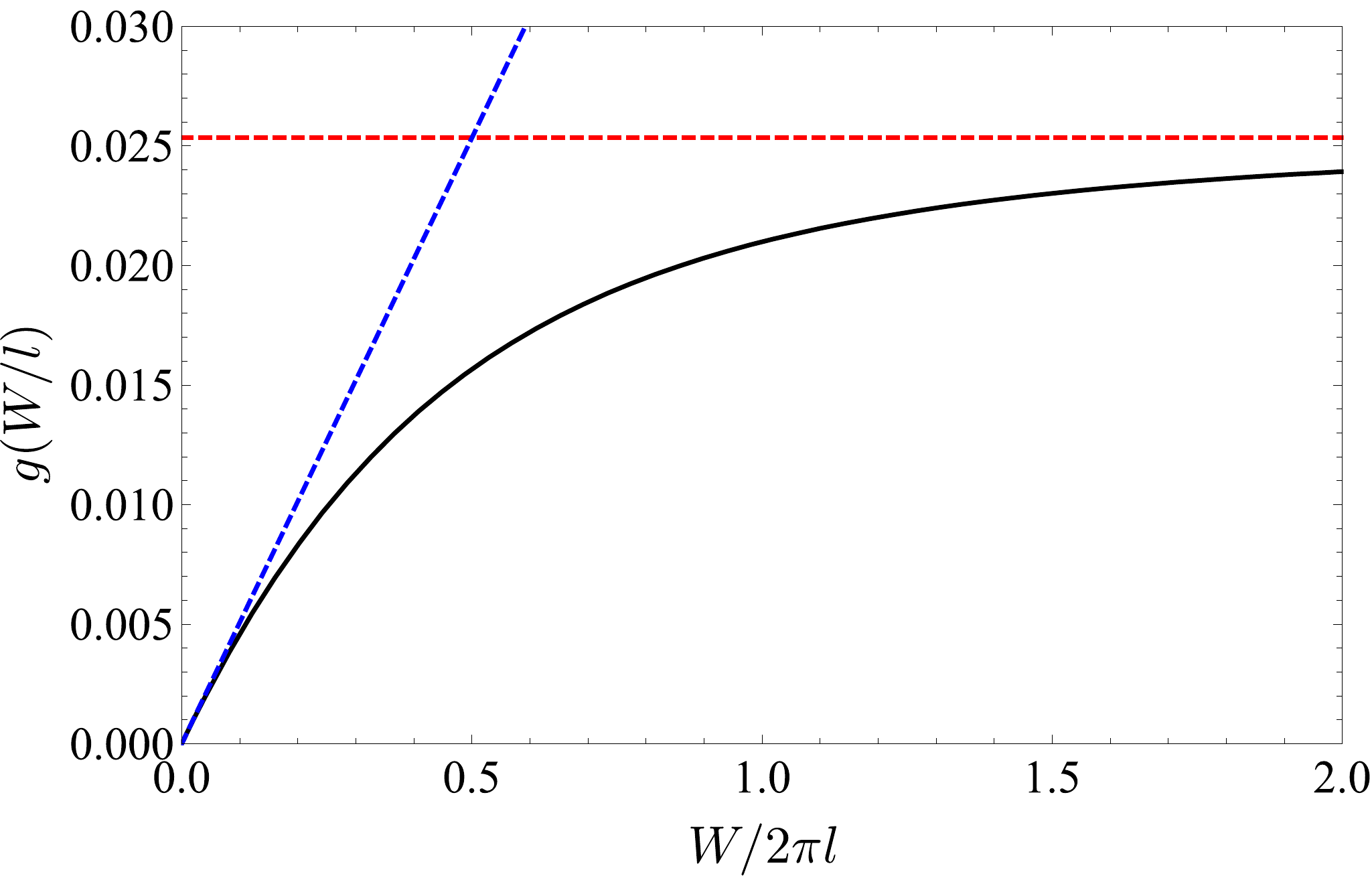}
\caption{The function $g(x)$ that determines the magnetic scattering rate in Eq.\ (\protect\ref{tby}) for a cylinder sample, Eq.\ (\protect\ref{gC}), with the circumference $W$. Dashed lines show the asymptotic behavior in the diffusive $W \gg l$ and ballistic $W \ll l$ limits.}
\label{gCf}
\end{figure}

In the case of a cylinder sample with circumference $W$ in transverse magnetic field, the vector potential is $A_x = (B W/2\pi) \cos(2\pi y/W)$. The corresponding phase factor is given by Eq.\ (\ref{alpha}),
\begin{equation}
 \label{alphacylinder}
 \alpha(y)
  = \frac{W^2}{2\pi^2 l_B^2}\, \sin(2\pi y/W).
\end{equation}
Substituting this result into Eq.\ (\ref{tBw}) and using Eq.\ (\ref{Rc}), we perform the integration over transverse coordinates $y$ and $y'$ and obtain the expression for $g(x)$ in the form of the angular integral
\begin{equation}
 \label{gC}
 g(x)
  = \int\limits_0^{2\pi} \frac{d\phi}{4\pi^3}\, \frac{x^2 \cos^2 \phi}{x^2 + 4 \pi^2 \sin^2 \phi}
  = \frac{x^2}{8 \pi^4} \left[
      \sqrt{\frac{4\pi^2}{x^2} + 1} - 1
    \right].
\end{equation}
This gives the following limiting values of the magnetic scattering rate (\ref{tby}) in diffusive and ballistic regimes:
\begin{equation}
 \label{tbc}
 \frac{1}{\tau_B}
  = \frac{W^2 l^2}{\tau l_B^4} \begin{dcases}
     1/4\pi^2, & W \gg l, \\
     W/4\pi^3 l, & W \ll l.
    \end{dcases}
\end{equation}
The former case reproduces Eq.\ (\ref{weakdiffuse}). The asymptotic behavior of $\tau_B$ agrees with the results obtained in Ref.\ \onlinecite{Takane}, where the prefactor in the ballistic limit $1/4\pi^3$ was computed approximately as $1/124$. The function $g(x)$ given by Eq.\ (\ref{gC}) is plotted in Fig.\ \ref{gCf} together with its two asymptotics.

\subsubsection{Flat sample}

\begin{figure}
\centering
\includegraphics[width = 0.9\columnwidth]{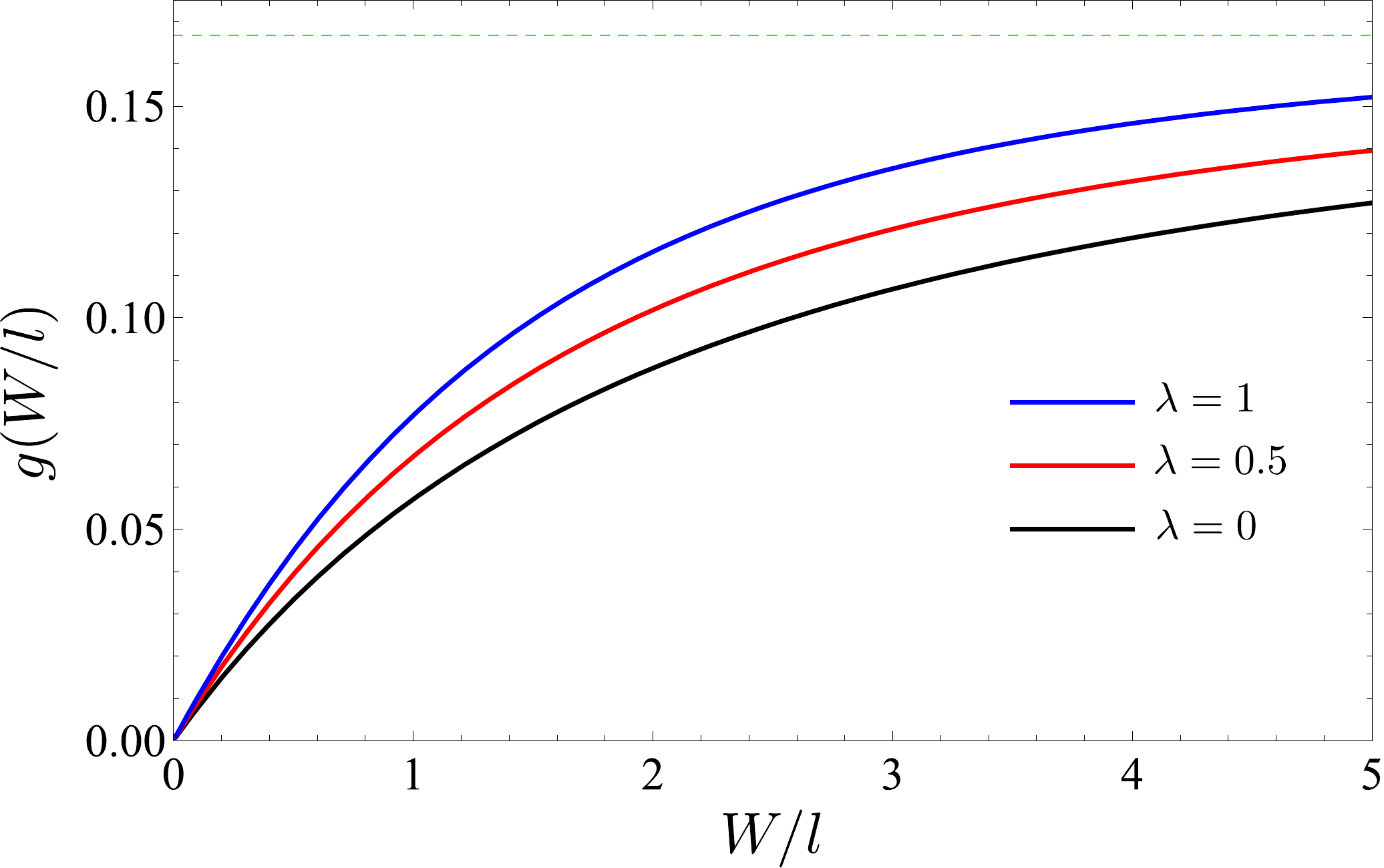}
\caption{The function $g(x)$ for a flat sample, Eq.\ (\protect\ref{gMD}), at different values of $\lambda$. In the diffusive limit $x = W/l \gg 1$, $g(x)$ attains the value $1/6$ shown by the dashed line. At $x \ll 1$, the function vanishes linearly according to Eq.\ (\protect\ref{cl}).}
\label{gMixedF}
\end{figure}

Let us now turn to the computation of the magnetic scattering rate for a flat sample. We use the gauge $A_x = B y$ which yields the phase factor (\ref{alpha})
\begin{equation}
 \label{alphaflat}
 \alpha(y)
  = l_B^{-2}\, (y^2 - W^2/4).
\end{equation}
This factor is to be substituted into Eq.\ (\ref{tBw}) together with the reduced Copperon from Eq.\ (\ref{Rf}). Similar to the calculation of the diffusion constant in the previous section, the $y$ and $y'$ dependence of the Cooperon is given by the exponential factor in Eq.\ (\ref{Ry}) and, hence, is easy to integrate. The angular integral over $\phi$ and $\phi'$ cancels the first term in Eq.\ (\ref{Rf}) since the factor $\cot\phi$ is odd under $\phi \mapsto \pi - \phi$. The rest of the Copperon involves the delta functions $\delta(\phi \pm \phi')$, hence, the final expression for the magnetic scattering rate can be written in terms of a single angular integral. The result can be represented in the form (\ref{tby}) with\footnote{In the two limiting cases of diffuse and mirror boundaries, an equivalent integral representation of this result was derived in Ref.\ \onlinecite{Gornyi}.}
\begin{multline}
 \label{gMD}
 g(x)
  = \frac{1}{6} - \frac{2 (1 - \lambda)}{3 \pi x} - \frac{\lambda}{2x^2} + \frac{16 (1 + \lambda)}{15 \pi x^3} \\
    -\int\limits_0^\pi \frac{d\phi}{\pi x^3}\, \cos^2 \phi \sin \phi
      \frac{\left[(1 - \lambda) x + 2 (1 + \lambda) \sin \phi \right]^2}{\lambda + e^{x/\sin \phi}}.
\end{multline}
This function is plotted in Fig.\ \ref{gMixedF} for several values of $\lambda$.

In the diffusive limit $W \gg l$, the parameter $x$ is large and the function $g(x)$ reduces to the constant $1/6$, which coincides with the result (\ref{weakdiffuse}). Other terms from the first line of Eq.\ (\ref{gMD}) provide small corrections to the magnetic scattering rate due to the sample boundaries. The integral term in the second line of Eq.\ (\ref{gMD}) is exponentially small. The situation is similar to the diffusion coefficient Eq.\ (\ref{Ddif}) obtained in the same limit.

\begin{figure}
\centering
\includegraphics[width = 0.9\columnwidth]{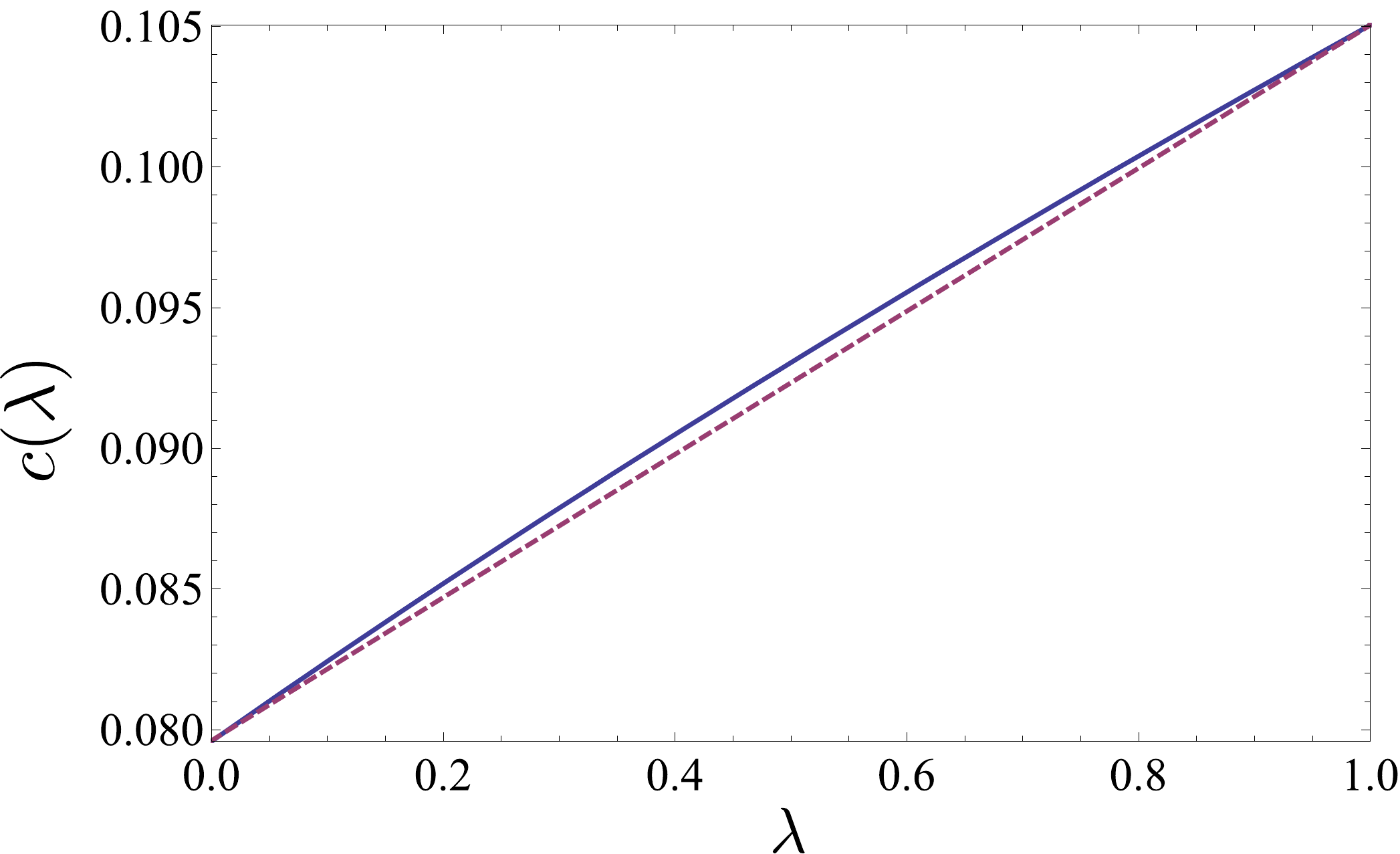}
\caption{The function $c(\lambda)$ [Eq.\ (\protect\ref{cl})] that determines the magnetic scattering rate in a flat ballistic sample in the weak field limit, Eq.\ (\protect\ref{tbf}). The two limiting values for diffuse ($\lambda = 0$) and mirror ($\lambda = 1$) boundaries are given by Eqs.\ (\protect\ref{climit}). We add the straight dashed line connecting the two endpoints to emphasize the small deviation of $c(\lambda)$ from the linear function.}
\label{cf}
\end{figure}

In the opposite case of a narrow sample $W \ll l$, the function $g(x)$ can be expanded in power series in the variable $x$. The first four terms of such an expansion in the integrand in Eq.\ (\ref{gMD}) exactly cancel the four terms in the first line of the same equation. The next term of this expansion is formally proportional to $x^2$ but multiplies a divergent angular integral. At the same time, numerical computation (cf.\ Fig.\ \ref{gMixedF}) suggests linear asymtotics at small $x$. To extract this linear term, we introduce the new integration variable $t = x/\sin\phi$ in Eq.\ (\ref{gMD}) and obtain an integral over the interval $t \in (x, \infty)$. By repeated partial integration, we eliminate all terms in the first line of Eq.\ (\ref{gMD}) and then take the limit of small $x$. This yields $g(x) = x c(\lambda)$ where the factor is given by
\begin{equation}
 \label{cl}
 c(\lambda)
  = \int\limits_0^\infty \frac{dt}{12\pi}\, \ln t\, \left(\frac{d}{dt} \right)^5 \frac{\left[(1 - \lambda) t + 2 (1 + \lambda) \right]^2}{\lambda + e^t}.
\end{equation}
This function is shown in Fig.\ \ref{cf}.

In the limits of diffuse and mirror boundaries, the factor $c(\lambda)$ equals
\begin{subequations}
\label{climit}
\begin{align}
 c(0)
  &= 1/4\pi, &&\text{(diffuse)}, \\
 c(1)
  &= 31 \zeta(5)/\pi^5, &&\text{(mirror)}.
\end{align}
\end{subequations}
These values agree with the results of Ref.\ \onlinecite{Beenakker} where $c(1)$ was computed approximately as $1/9.5$. The magnetic scattering rate is given
by Eq.\ (\ref{tby}),
\begin{equation}
 \label{tbf}
 \frac{1}{\tau_B}
  = \frac{W^3 l}{\tau l_B^4}\, c(\lambda),
 \qquad
 W \ll l.
\end{equation}

\begin{figure}
\centering
Diffusive limit $W \gg l$
\begin{equation*}
\vcenter{\hbox{\includegraphics[width = 0.2 \textwidth]{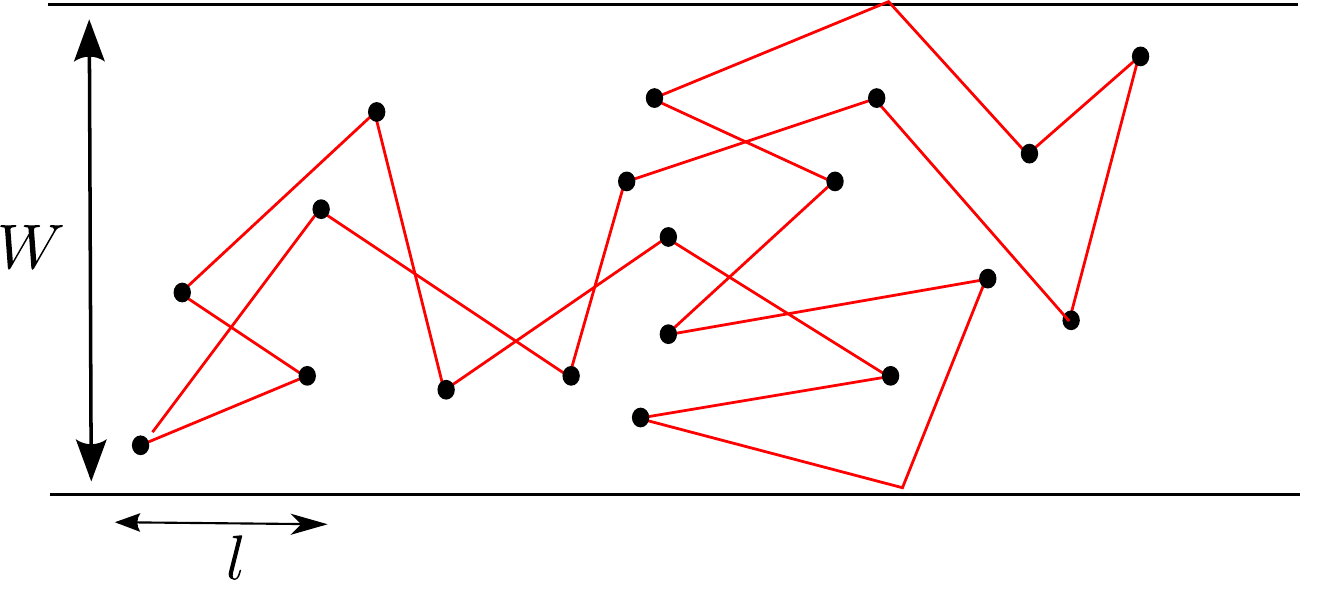}}} \Rightarrow l_B^2 = W \sqrt{D \tau_B} 
\end{equation*}
Ballistic limit $W \ll l$, weak field $l_B \gg \sqrt{W l}$
\begin{eqnarray*}
\vcenter{\hbox{\includegraphics[width = 0.2 \textwidth]{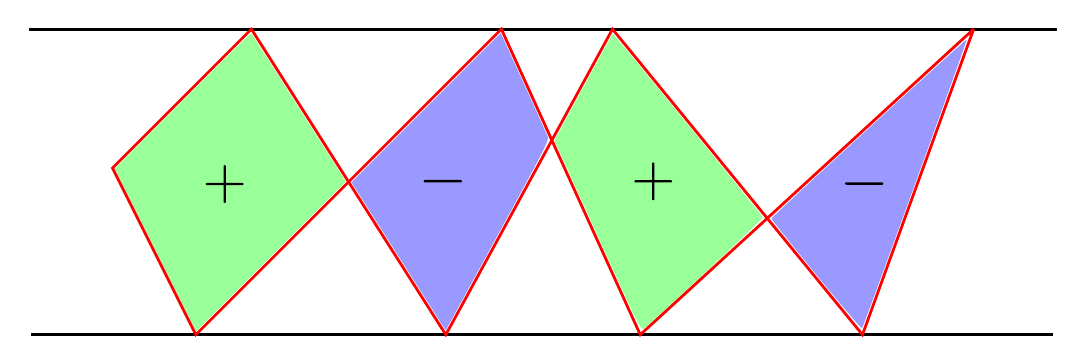}}} && \phi \gtrsim W/l \rightarrow \text{flux cancellation}
\end{eqnarray*}
\caption{Schematic illustration of the trajectories providing the major contribution to the magnetic scattering rate in the diffusive and ballistic regimes.}
\label{traj}
\end{figure}

The different dependence of the magnetic scattering rate $\tau_B^{-1}$ in the diffusive and ballistic regimes can be qualitatively understood as follows. In the diffusive limit, $W \gg l$, electron mobility is mainly governed by the bulk scattering. The effect of the boundary is only to restrict the spatial extent of the diffusion in the $y$ direction as shown in figure \ref{traj}. To estimate the magnetic scattering time, we notice that a diffusive trajectory returning to the origin in time $\tau_B$ covers an area $\sim W \sqrt{D \tau_B}$. The magnetic scattering time is determined by comparing this area to the area $l_B^2$ pierced by a single flux quantum. This yields the estimate $1/\tau_B \sim W^2 D/l_B^4$ in accordance with Eq.\ (\ref{weakdiffuse}) and Ref.\ \onlinecite{Altshuler81}.

In the opposite ballistic limit $W \ll l$, the particle scatters predominantly at the edge of the sample. This leads to the effect of flux cancellation first pointed out in Ref.\ \onlinecite{Dugaev}. Any closed trajectory that does not experience bulk scattering has exactly zero area and is hence insensitive to the magnetic field, see Fig. \ref{traj}. Trajectories contributing to magnetoresistance have many bulk scatterings but their area is effectively reduced due to this geometric cancellation. Namely, a typical area can be estimated as $W^2$ per each bulk scattering. However, as can be seen from Eqs.\ (\ref{gC}) and (\ref{gMD}), the main contribution comes from the shallow trajectories with angles $|\sin \phi| \sim W/l \ll 1$. Such trajectories effectively avoid flux cancellation but their probability has an extra small factor $W/l$. This factor appears in the magnetic scattering rate yielding $1/\tau_B \sim W^3 l/\tau l_B^4$ in accordance with Eqs.\ (\ref{tbc}) and (\ref{tbf}).

We can estimate the number of impurities involved in a typical trajectory contributing to magnetoconductivity of a narrow sample $W \ll l$. Within the time $\tau_B$, the number of bulk scattering events is $\tau_B/\tau \sim l_B^4/W^3 l \gg 1$. The phase accumulated between two consecutive bulk scatterings for a typical shallow trajectory is of the order $Wl/l_B^2$. Throughout this Subsection we assumed that this phase is small, which allowed us to expand exponential factors in Eq.\ (\ref{tauB}) and use the simplified expression (\ref{tBw}). From the above estimate we see that such an approximation is justified for a relatively weak magnetic fields $l_B \gg \sqrt{Wl}$.

\subsection{Magnetic scattering rate in a strong magnetic field $W \ll l_B \ll \sqrt{W l}$}
\label{Sec:strong}

In a ballistic sample, $W \ll l$, a qualitatively different regime of strong magnetic field is possible. When the magnetic length $l_B$ drops below $\sqrt{Wl}$, the phase factors acquired with each bulk scattering are no longer small. At the same time, the quasi-1D dynamics is still diffusive and the magnetoresistance is determined by long trajectories involving many impurities provided $l_B \gg W$. In this limit, the magnetic field cannot be treated as a small perturbation and we should use the general expression (\ref{tauB}) for the magnetic scattering rate.

Angular integrals in Eq.\ (\ref{tauB}) involve two types of exponential factors. First, the reduced Cooperon propagator, Eqs.\ (\ref{Rf}) and (\ref{Rc}), contains the terms $\sim \exp(-W/l|\sin\phi|)$. Due to these terms, $W/l$ acts as an effective lower cutoff for the angular integration in the calculation of the diffusion coefficient and magnetic scattering rate in a weak field. Magnetic phase $\alpha$ [see Eq.\ (\ref{alpha})] yields another exponential factor $\sim \exp[i(W/l_B)^2 \cot\phi]$. In the strong field limit $l_B \ll \sqrt{Wl}$, this latter factor becomes more important and effectively replaces the lower cutoff for the angular integral by $(W/l_B)^2$. This allows us to simplify the calculation of the magnetic scattering rate by taking the formal limit $W/l \to 0$ while keeping $(W/l_B)^2$ small but nonzero.

\subsubsection{Cylinder sample}

The reduced Cooperon propagator (\ref{Rc}) for a cylinder sample takes a very simple form in the limit $W/l \to 0$. Neglecting the width of the sample we also completely disregard the dependence on $y$ and $y'$ thus having $\mathcal{R} = 2\pi \delta(\phi - \phi')$. With such a simple expression for the Cooperon, we readily perform the angular integration in Eq.\ (\ref{tauB}) and obtain
\begin{equation}
 \label{tb1}
 \frac{1}{\tau_B}
  = \int \frac{dy\, dy'}{\tau W^2} \left[
      1 - e^{ -\left| \alpha(y) - \alpha(y') \right| }
    \right].
\end{equation}
Here the magnetic phase $\alpha(y)$ is given by Eq.\ (\ref{alphacylinder}). As was pointed out before, these phase factors are of the order of $(W/l_B)^2 \ll 1$ and provide an effective lower cutoff for the angular integral. As a result, the expression (\ref{tb1}) is non-analytic in $\alpha$. However, once angles are integrated out, it is safe to expand Eq.\ (\ref{tb1})  in small $\alpha$. The remaining integral over $y$ and $y'$ yields the magnetic scattering rate
\begin{equation}
 \frac{1}{\tau_B}
  = \int \frac{dy\, dy'}{2\pi^2 \tau l_B^2}
      \left| \sin\frac{2\pi y}{W} - \sin\frac{2\pi y'}{W} \right|
    = \frac{4 W^2}{\pi^4 \tau l_B^2}.
\end{equation}
This expression agrees with the result of Ref.\ \onlinecite{Takane}. Unlike the case of the weak magnetic field, the rate $1/\tau_B$ is linear in magnetic field.

\subsubsection{Flat sample}

In the case of a flat sample, we will first consider the two limits of mirror and diffuse boundaries and then derive a general expression for arbitrary $\lambda$. For specular boundaries, the calculation is very similar to the case of a cylinder sample discussed above. Taking the limit $W/l \to 0$ in Eq.\ (\ref{Rf}) and setting $\lambda = 1$, we obtain the reduced Cooperon propagator $\mathcal{R} = \pi [\delta(\phi + \phi') + \delta(\phi - \phi')]$. After angular integration in Eq.\ (\ref{tauB}), we obtain
\begin{equation}
 \frac{1}{\tau_B}
  = \frac{1}{\tau} \left[
      1 - \int \frac{dy\, dy'}{2 W^2} \left(
        e^{ -\left| \alpha - \alpha' \right| } + e^{ -\left| \alpha + \alpha' \right| }
      \right)
    \right].
\end{equation}
Here, we use the short notations $\alpha = \alpha(y)$ and $\alpha' = \alpha(y')$ for the magnetic phase given by Eq.\ (\ref{alphaflat}). Since this phase contains a small factor $(W/l_B)^2$, we expand the exponential terms and obtain the result
\begin{equation}
 \label{tBmirror}
 \frac{1}{\tau_B}
  = \int \frac{dy\, dy'}{2 \tau W^2} \bigl(
       \left| \alpha - \alpha' \right| + \left| \alpha + \alpha' \right|
      \bigr)
  = \frac{5 W^2}{24\tau l_B^2}.
\end{equation}
This value of the magnetic scattering rate reproduces the result of Ref.\ \onlinecite{Beenakker}.

Consider now the opposite case of diffuse boundary, $\lambda = 0$. The Cooperon propagator (\ref{Rf}) reduces to $\mathcal{R} = 1$ in the limit $W/l \to 0$. This means that both the angles $\phi$, $\phi'$ and the coordinates $y$, $y'$ are completely decoupled in Eq.\ (\ref{tauB}). Magnetic scattering rate thus takes the form
\begin{equation}
 \frac{1}{\tau_B}
  = \frac{1 - \langle\!\langle e^{i \alpha \cot\phi} \rangle\!\rangle^2}{\tau}
  = \frac{1}{\tau} \left[
      1 - \left( \int \frac{dy}{W}\, e^{-|\alpha|} \right)^2
    \right].
\end{equation}
Now we expand the integrand in small $\alpha$. Since we are interested in the magnetic scattering rate in the leading order in $(W/l_B)^2$, it suffices to expand only one of the two $y$ integrals and replace the other integral by $1$. The result reads
\begin{equation}
 \label{tBdiffuse}
 \frac{1}{\tau_B}
  = \frac{2}{\tau} \int \frac{dy}{W}\, |\alpha|
  = \frac{W^2}{3 \tau l_B^2}.
\end{equation}
This expression also agrees with the result of Ref.\ \onlinecite{Beenakker}.

\begin{figure}
\centering
\includegraphics[width = 0.45 \textwidth]{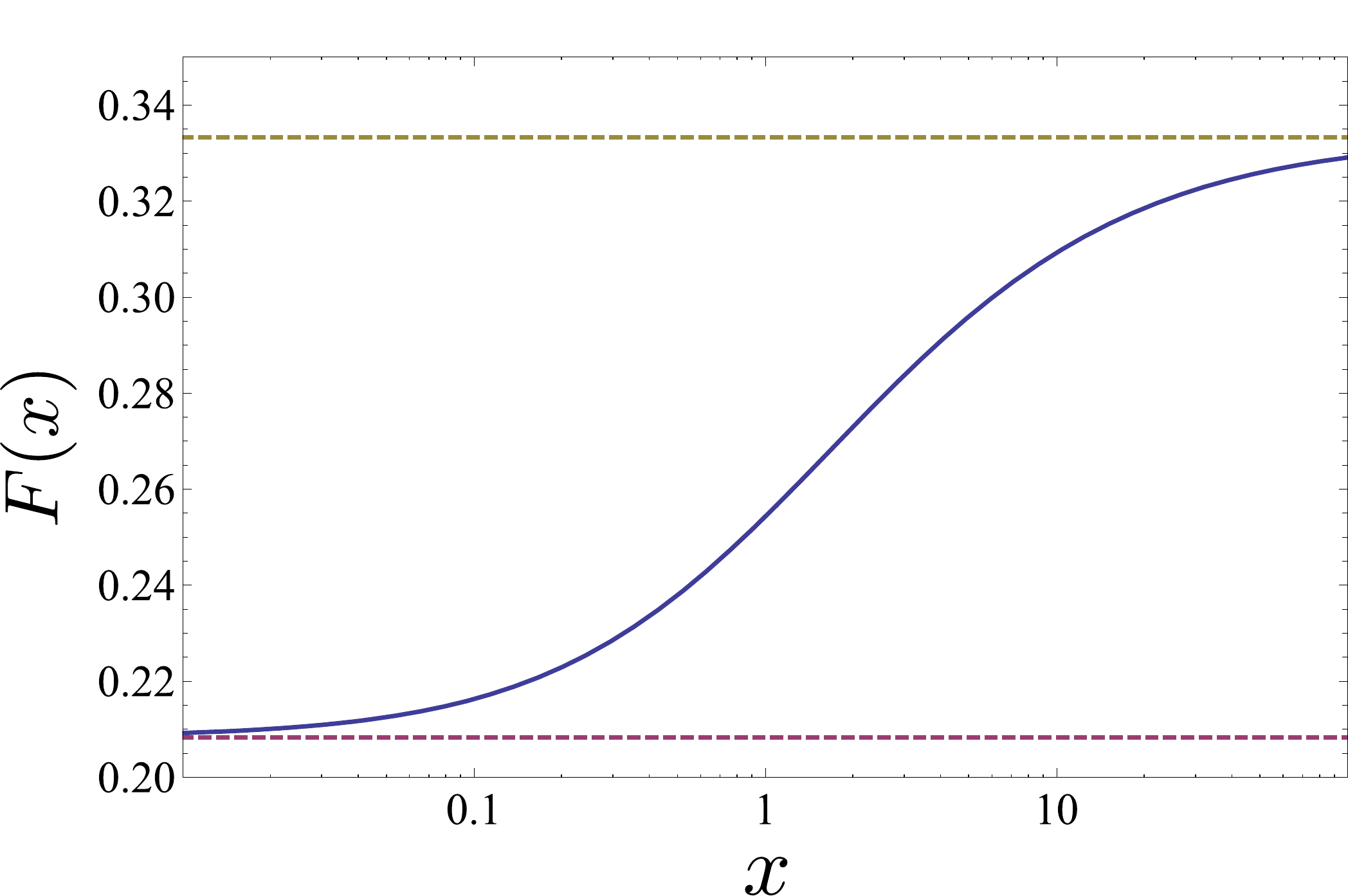}
\caption{Crossover function $F(x)$ defined in Eq.\ (\protect\ref{Fx}) for the magnetic scattering rate in the flat ballistic sample. Dashed lines show the limiting values for mirror ($x = 0$) and diffuse ($x = \infty$) boundaries.}
\label{FF}
\end{figure}

Let us now consider the general case of an arbitrary boundary. Our goal is to find an expression that interpolates between Eqs.\ (\ref{tBmirror}) and (\ref{tBdiffuse}) as $\lambda$ changes from $1$ to $0$. The situation is similar to the computation of the diffusion coefficient (\ref{Dphi}) in the narrow sample. The crossover from mirror to diffuse boundary should happen at some value of $\lambda$ close to $1$ such that the probability of bulk scattering is comparable to the probability of diffuse scattering at the boundary. We will take the limit $W/l \to 0$ and simultaneously the limit $\lambda \to 1$ assuming that the parameter $b$, defined in Eq.\ (\ref{b}), remains fixed. This yields the following expression for the reduced Cooperon (\ref{Rf}):
\begin{gather}
 \begin{multlined}[b][0.8\columnwidth]
  \mathcal{R}
   = \pi \biggl[
       \frac{b\, |\sin\phi \sin\phi'|}{2\Gamma_- (b |\sin\phi| + 1)(b |\sin\phi'| + 1)} \\
       +\frac{\delta(\phi - \phi') + \delta(\phi + \phi')}{b |\sin\phi| + 1}
     \biggr],
 \end{multlined} \label{Rb} \\
 \Gamma_-
  = \frac{1}{2} \int_0^\pi \frac{d\phi\, \sin\phi}{b \sin\phi + 1}.
\end{gather}
The two terms in the propagator $\mathcal{R}$ have different nature. In the first line of Eq.\ (\ref{Rb}), the angles $\phi$ and $\phi'$ are decoupled as in the case of diffuse boundary while the second line contains delta functions analougous to the mirror boundary limit. Upon substitution of Eq.\ (\ref{Rb}) into Eq.\ (\ref{tauB}), we obtain the magnetic scattering rate in the form
\begin{multline}
 \frac{1}{\tau_B}
  = \frac{1}{\tau} \biggl\{
    1 - \frac{\pi b}{2\Gamma_-} \left[\int \frac{dy}{W} \int_0^\pi \frac{d\phi\, \sin\phi\, \cos(\alpha \cot\phi)}{\pi (b \sin\phi + 1)} \right]^2 \\
    - \int_0^\pi \frac{d\phi}{\pi(b \sin\phi + 1)} \left[ \int \frac{dy}{W}\, \cos(\alpha \cot\phi) \right]^2 \biggr\}.
\end{multline}
We further symplify this expression using the small parameter $(W/l_B)^2$. The second term in curly braces contains two identical integrals that involve $\cos(\alpha \cot\phi) \approx 1$ for not very small angles $\phi$. We expand this term up to the linear order in $\cos(\alpha \cot\phi)$ at the point $1$. This leads to the cancellation of $\Gamma_-$ in the denominator and allows us to combine the terms in a single angular integral
\begin{multline}
 \frac{1}{\tau_B}
  = \frac{4}{\tau} \int_0^\pi \frac{d\phi}{\pi} \int \frac{dy}{W} \sin^2\left(\frac{\alpha}{2} \cot\phi\right) \\
    \times \left[
      1 - \int \frac{dy'}{W} \frac{\sin^2\bigl(\frac{\alpha'}{2} \cot\phi\bigr)}{b \sin\phi + 1}
    \right].
\end{multline}
Since the phases $\alpha$ and $\alpha'$ are small, the main contribution to the angular integral comes from $\phi$ close to $0$ or $\pi$. We can take advantage of this fact by introducing the new integration variable $t = 8 (W/l_B)^2 \cot\phi$. Using Eq.\ (\ref{alphaflat}) and rescaling $y$ by $W$, we finally represent the magnetic scattering rate in terms of a single parameter crossover function
\begin{gather}
\label{tbStrong}
 \frac{1}{\tau_B}
  = \frac{W^2}{\tau l_B^2} F\left( (1-\lambda)\frac{W l}{8 l_B^2} \right), \\
\label{Fx}
 F(x)
  = \frac{1}{3} - \int_0^\infty\!\!\! \frac{dt}{\pi t (x + t)} \left( \int_0^1 dy \sin^2 \bigl[ t (1 - y^2) \bigr] \right)^2.
\end{gather}
This function is plotted in Fig.\ \ref{FF}.

Typical trajectories contributing to magnetoresistance in the strong field limit have the angle $\phi \lesssim (W/l_B)^2$. Such trajectories scatter $\sim W l/l_B^2 \gg 1$ times at the boundary between two successive bulk scatterings. This explains the strong sensitivity of $\tau_B$ to the boundary roughness $1 - \lambda$ and the origin of the crossover parameter $x$ in Eq.\ (\ref{tbStrong}).

Remarkably, the crossover from the mirror to the diffuse boundary limit is governed by the parameter $x$ which depends on magnetic field. For a narrow sample with almost specular boundaries (such that $W/l \ll 1 - \lambda \ll 1$), magnetic scattering rate will gradually change its behavior from Eq.\ (\ref{tbf}) to Eq.\ (\ref{tBmirror}) and further to Eq.\ (\ref{tBdiffuse}) with increasing magnetic field. Moreover, in the situation $W/l \ll 1 - \lambda \ll l_B^2/Wl \ll 1$, magnetic scattering rate is given by Eq.\ (\ref{tBmirror}) corresponding to mirror edges, while the diffusion constant [first case of Eq.\ (\ref{Dcross})] is dominated by the randomized boundary scattering. This resolves the discrepancy observed in some previous works.\cite{Katine, Niimi}

\subsection{Magnetic scattering rate $\tau_B^{-1}$ in the general case}
\label{Sec:q1d}

\begin{figure}
\centering
\includegraphics[width = 0.45 \textwidth]{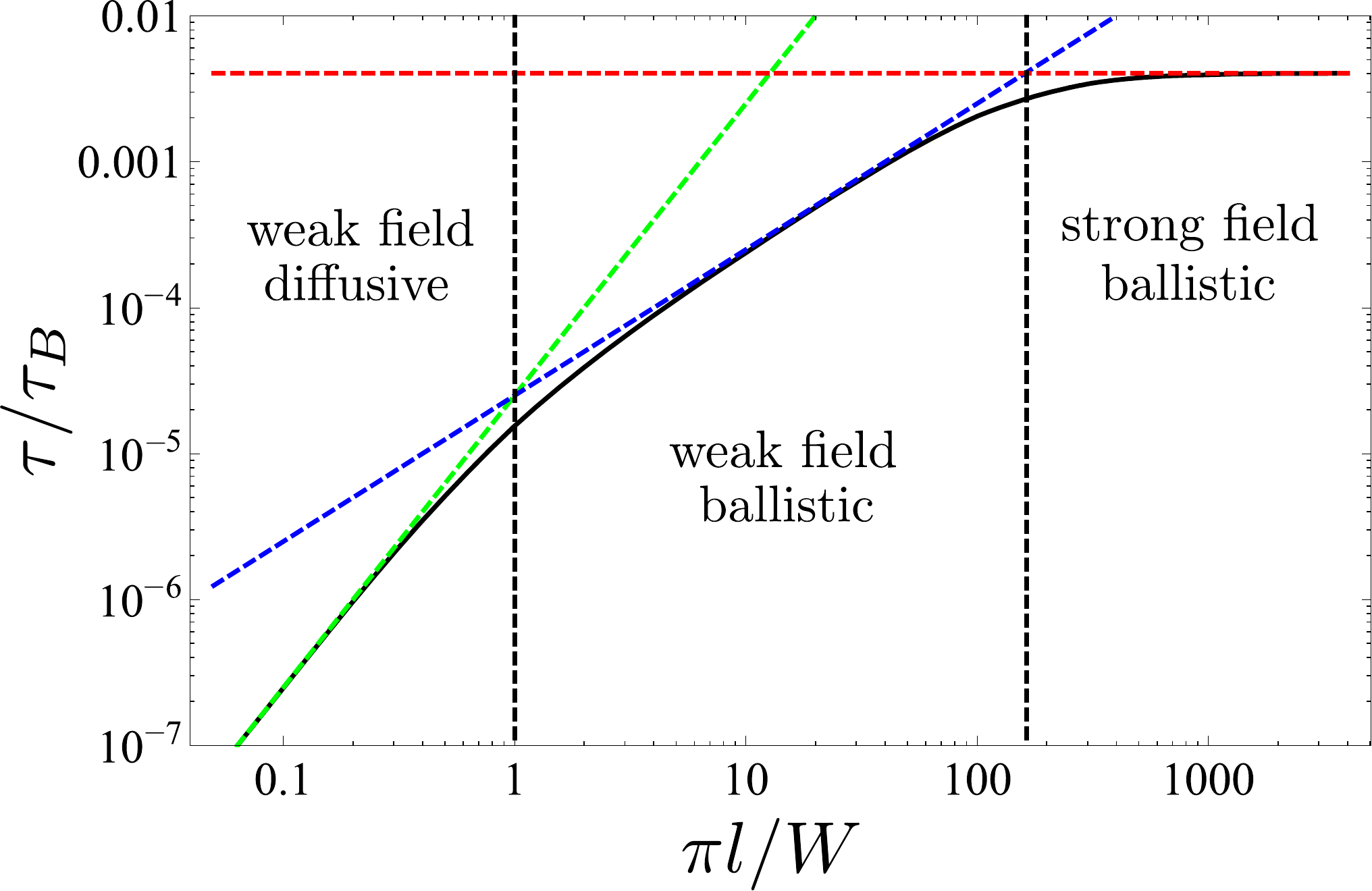}
\caption{Magnetic scattering rate in units of $1/\tau$ as a function of the mean free path $l$ (in units of $W/\pi$) for a cylinder sample with $\pi l_B/W = 10$. The crossover from weak field diffusive to weak field ballistic and further to strong field ballistic limits occurs as $l$ is increased.}
\label{tCf}
\end{figure}

In the previous Sections, we have studied magnetic scattering rate in the two crossover regimes: between ballistic and diffusive limit ($W \sim l \ll l_B$) or between weak and strong magnetic field in the ballistic case ($W \ll \sqrt{W l} \sim l_B$). Here we will consider a general situation where, apart from the quasi-1D condition $l_B \ll W$, the parameters $l$, $W$ and $l_B$ are arbitrary. In the general case, magnetic scattering rate is given by Eq.\ (\ref{tauB}) with the Cooperon propagator Eq.\ (\ref{Rc}) for a cylinder sample or Eq.\ (\ref{Rf}) for a flat sample.

\begin{figure}
\centering
\includegraphics[width = 0.45 \textwidth]{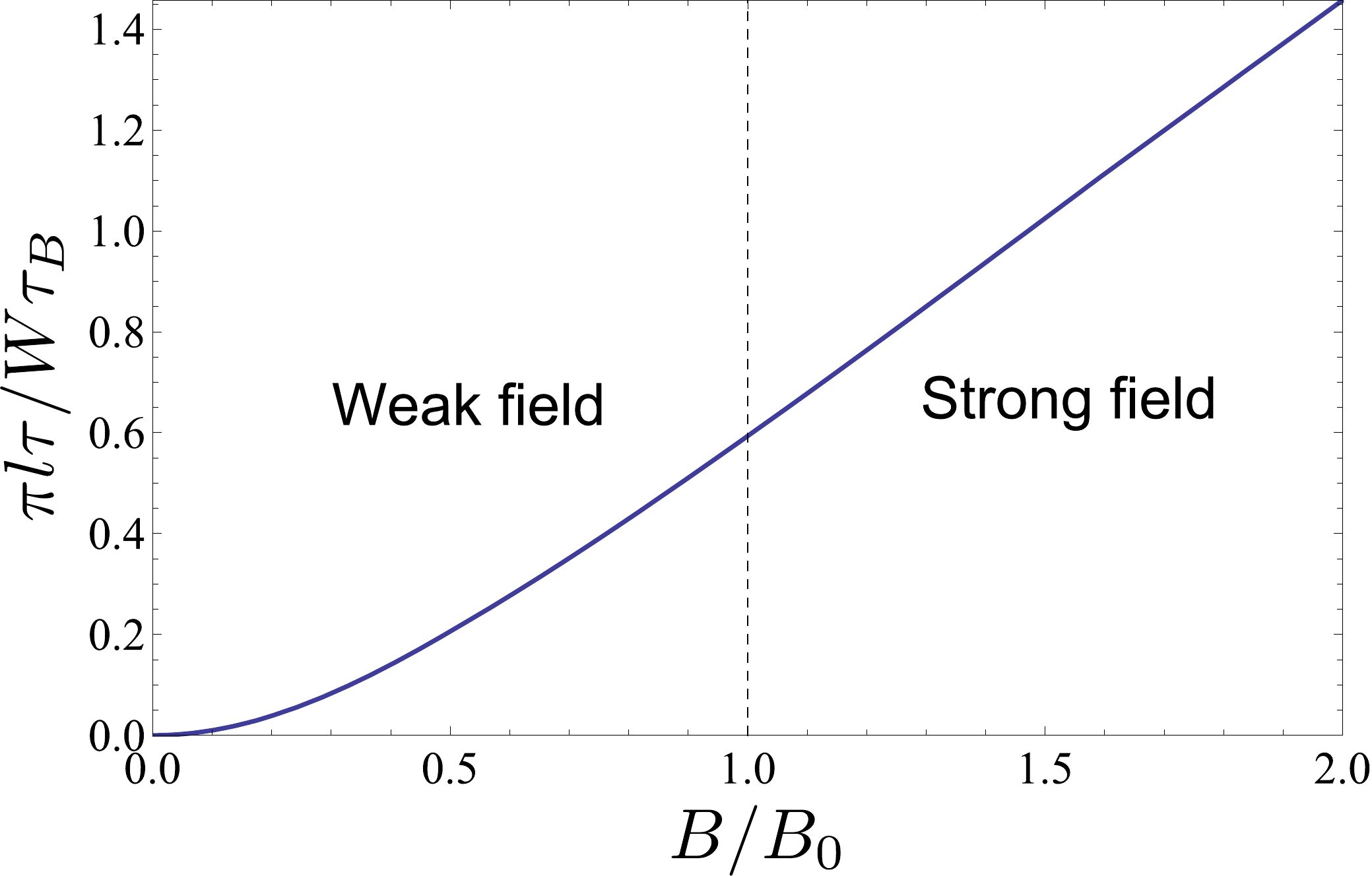}
\caption{Magnetic scattering rate (in units of $W/\pi l \tau$) as a function of the magnetic field $B$ (in units of $B_0 = h/e W l$) for a cylinder sample with fixed $W/2\pi l = 10^{-3}$. The crossover from weak (quadratic) to strong (linear) field behavior occurs at $B \sim B_0$.}
\label{tCylBF}
\end{figure}

For cylinder geometry, the result contains two spatial integrals over $y$ and $y'$ and a single angular integral. Such expression can hardly be further simplified without extra assumptions about the values of $l$, $W$ and $l_B$. We calculate the magnetic scattering rate in this case numerically and plot the result in Fig.\ \ref{tCf} as a function of disorder strength (quantified by the mean free path $l$) at fixed values of the magnetic field and sample width. As $l$ is increased, crossovers between the three regimes of diffusive transport in weak field, ballistic transport in weak field, and ballistic transport in strong field can be observed. An alternative setting is considered in Fig.\ \ref{tCylBF} where the magnetic scattering rate is plotted as a function of magnetic field for fixed width and disorder strength. The dependence of the magnetic scattering rate on the magnetic field changes from quadratic ($\tau_B^{-1} \propto B^2$) to linear ($\tau_B^{-1} \propto B$) as the ballistic system undergoes the crossover from weak to strong field limit.

\begin{figure}
\centering
\includegraphics[width = 0.45 \textwidth]{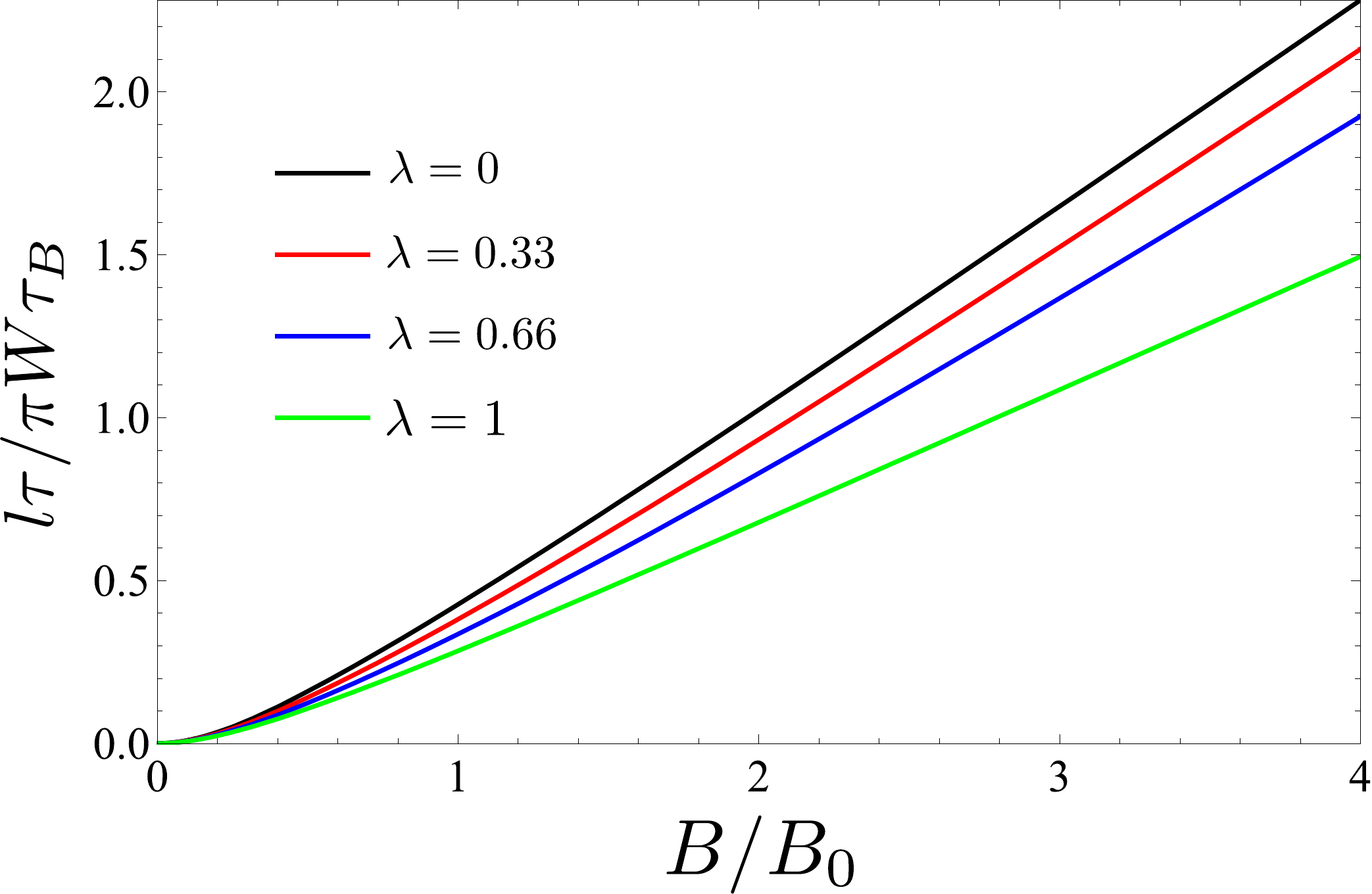}
\caption{Magnetic scattering rate (in units of $\pi W/l \tau$) as a function of the magnetic field $B$ (in units of $B_0 = h/e W l$) for a flat sample with fixed $W/l = 10^{-3}$ at different values of boundary roughness parameter $\lambda$.}
\label{tMixedBF}
\end{figure}

In the case of the flat sample, the expression for the magnetic scattering rate (\ref{tauB}) with the Cooperon propagator (\ref{Rf}) involves two spatial and two angular integrals. Numerical computation of these integrals yields the result plotted in Fig.\ \ref{tMixedBF}. Similarly to the cylinder case (cf.\ Fig.\ \ref{tCylBF}), the system demonstrates a crossover from weak to strong field limit with quadratic and linear dependence of the magnetic scattering rate, respectively. The slope of the linear dependence in strong fields changes between $5/24$ and $1/3$ depending on the value of the parameter $\lambda$, see Eq.\ (\ref{tbStrong}).

\section{Quasi-two-dimensional case}
\label{Sec:q2d}

Let us briefly discuss an extension of our results to the case of a quasi-2D system in a parallel magnetic field. The sample has the geometry of a thin film with the thickness $W$ much smaller than the magnetic length $l_B$. This situation was considered first in Ref.\ \onlinecite{Altshuler81} in the diffusive limit ($W \gg l$) and later in Refs.\ \onlinecite{Dugaev, Beenakker} also in the ballistic system with diffuse or mirror boundaries. We will now present the results for a general boundary with the specularity parameter $\lambda$.

In order to keep notations coherent, we consider a thin film with width $W$ in the $y$ direction extended along $x$ and $z$. We assume a spherical Fermi surface and denote the velocity direction by the unit vectors $\mathbf{n}$. The boundary conditions for a general case generalize Eq.\ (\ref{mixed}) and have the form
\begin{multline}
 \mathcal{C}(y = \pm W/2, \mathbf{n}_\mp)
  = \lambda\, \mathcal{C}(y = \pm W/2, \mathbf{n}_\pm) \\
    + 4(1 - \lambda) \int_{n'_y > 0} \!\! d\mathbf{n}' \; n'_y \; \mathcal{C}(y = \pm W/2, \mathbf{n}'_\pm).
\end{multline}
Here $\mathbf{n}$ denotes a unit vector with $n_y > 0$ and the integration measure $d\mathbf{n}'$ is normalized such that the integral over the full sphere is unity. The notation $\mathbf{n}_\pm$ denotes the vector $(n_x, \pm n_y, n_z)$. We choose again the Landau gauge such that $\mathbf{A} = (B y, 0, 0)$. The weak localization correction is given by the Cooperon at coincident points averaged over space and angular directions, see Eq.\ (\ref{Ds}). Following the same argument presented in Sec.\ \ref{Sec:formalism}, we can show that only the angular and spatial average of the reduced Cooperon $\mathcal{C}_0$ defined in Eq.\ (\ref{C0}) is needed for computing the magnetoconductance. We can also single out the dependence on the magnetic field similar to Eq.\ (\ref{C0R}) and expand the Cooperon in the small 2D momentum $\mathbf{q}$ in the $xz$ plane. The result for magnetoconductivity reads
\begin{equation}
 \label{Ds2D}
 \Delta \sigma_{2D}(B)
  = -\eta\, \frac{e^2}{2\pi h} \ln \left(1 + \frac{\tau_{\phi}}{\tau_B} \right).
\end{equation}
The main difference from the quasi-1D case Eq.\ (\ref{MR}) is the logarithmic dependence on $\tau_B$. The diffusion coefficient and the magnetic scattering rate are given by
\begin{align}
 \label{D2}
  D &= v_0 l \langle\!\langle n_x \mathcal{R} \; n_x \rangle\!\rangle = v_0 l \langle\!\langle n_z \mathcal{R} \; n_z \rangle\!\rangle, \\
 \label{tauB2}
  \frac{1}{\tau_B} &= \frac{1 - \langle\!\langle e^{i (n_x/n_y) \alpha} \mathcal{R} \; e^{-i (n_x/n_y) \alpha} \rangle\!\rangle}{\tau},
\end{align}
with $\alpha$ defined in Eq.\ (\ref{alpha}).

A general expression for the reduced Cooperon $\mathcal{R}$ can be obtained following steps similar to Appendix \ref{App:Cooperon}. The final result coincides with Eq.\ (\ref{Rf}) with $\sin \phi$ replaced everywhere by $n_y$, $2 \pi \delta(\phi \pm \phi')$ replaced by $\delta(\mathbf{n} - \mathbf{n}'_\pm)$ and the prefactor of the $(1 - \lambda)$ term changed from $\pi/2$ to $2$ (which is $1/\langle |n_y| \rangle$). The results for the diffusion coefficient and magnetic scattering rate are obtained using Eqs.\ (\ref{D2}) and (\ref{tauB2}) using manipulations similar to those used in the quasi-1D case.

The diffusion constant is given by the following integral [cf.\ Eq.\ (\ref{Dm})]:
\begin{multline}
\label{Dq2D}
 D
  = \frac{v_0 l}{3} \left[ \rule{0pt}{7mm} 1 - (1 - \lambda) \frac{3 l}{2 W} \right. \\ \left.
      \times \int_0^{\pi/2} \! \! d\phi\, \sin \phi\, \cos^3 \phi\,
      \frac{1 - e^{-\tfrac{W}{l \sin \phi}}}{1 - \lambda e^{-\tfrac{W}{l \sin \phi}}}
    \right].
\end{multline}
It reproduces the known value $v_0 l/3$ in the diffusive limit $W \gg l$ or in a system with the mirror boundary ($\lambda = 1$) and the result of Ref.\ \onlinecite{Beenakker} for a sample with rough boundaries ($\lambda = 0$). In the ballistic limit $W \ll l$, we introduce the parameter $b$ defined by Eq.\ (\ref{b}) and derive the result [cf.\ Eq.\ (\ref{Dphi})]
\begin{align}
 \label{Db2D}
 D
  &= \frac{v_0 l}{3} \int_0^{\pi/2} \frac{d\phi\, \cos^3 \phi}{1 + b \sin \phi} \nonumber \\
  &= \frac{v_0 l}{2 b^3} \left[
      1 - \frac{b^2}{2} + (b^2 - 1) \log(1 + b)
    \right].
\end{align}
This function describes the crossover between the regimes of predominantly bulk ($b \ll 1$) or boundary ($b \gg 1$) scattering with the limiting values [cf.\ Eq.\ (\ref{Dcross})]
\begin{equation}
 D
  = v_0 l \begin{dcases}
      \frac{(1 + \lambda)W}{4 (1 - \lambda) l} \ln \frac{(1 - \lambda)l}{(1 + \lambda)W}, & W/l \ll 1 - \lambda, \\
      \frac{1}{3} - \frac{(1 - \lambda)l}{8 W}, & 1 - \lambda \ll W/l \ll 1.
    \end{dcases}
\end{equation}

In a weak parallel magnetic field, $l_B \gg \sqrt{W l}, W$, the magnetic scattering rate is given by Eq.\ (\ref{tby}) with the function $g(x)$ of the form
\begin{multline}
 \label{g2D}
 g(x)
  = \frac{1}{9} - \frac{1 - \lambda}{8 x}  - \frac{4 \lambda}{15 x^2}  + \frac{1 + \lambda}{6 x^3} \\
    -\int\limits_0^{\pi/2} \frac{d\phi}{2 x^3}\, \cos^3 \phi \sin \phi \frac{\left[(1 - \lambda) x + 2 (1 + \lambda) \sin \phi \right]^2}{\lambda + e^{x/\sin \phi}}.
\end{multline}
For a thick diffusive film, $x = W/l \gg 1$, this function attains the constant value $1/9$ in agreement with Refs.\ \onlinecite{Altshuler81, Dugaev}. In the opposite ballistic limit $x \ll 1$, the result depends on the type of the boundary [cf.\ Eq.\ (\ref{tbf})] with $g(x) = (\pi x/4) c(\lambda)$. Here $c(\lambda)$ is exactly the same universal crossover function, defined by Eq.\ (\ref{cl}) and shown in Fig.\ \ref{cf}, as was found in the quasi-1D case.

In the strong field limit, $W \ll l_B \ll \sqrt{W l}$, the magnetic scattering rate is given by
\begin{equation}
 \label{tbStrong2D}
 \frac{1}{\tau_B}
  = \frac{W^2}{2\tau l_B^2} \int_0^\pi \! \! d \theta \; \sin \theta \; F\left( (1-\lambda)\frac{W l}{8 l_B^2} \sin \theta \right),
\end{equation}
with the function $F(x)$ defined by Eq.\ (\ref{Fx}). We thus conclude that $\tau_B^{-1}$ is proportional to $W^2/\tau l_B^2$ with the prefactor changing from $5/24$ in the limit $1 - \lambda \ll l_B^2/Wl \ll 1$ (mirror boundaries) to $1/3$ for $1 - \lambda \gg l_B^2/Wl$ (diffuse boundaries). This dependence is qualitatively similar to the quasi-1D case.

\section{Summary and discussion}
\label{Sec:discussion}

\begin{table*}
\center
\begin{tabular}{lcccccc}
\hline\hline
\multirow{3}{*}{Limiting case} & \multirow{3}{*}{Quantity} & \multicolumn{3}{c}{quasi-1D} & \multicolumn{2}{c}{quasi-2D} \\
\cline{3-7}
& & \strut\enskip\multirow{2}{*}{cylinder}\enskip\strut & \multicolumn{2}{c}{flat channel} & \multicolumn{2}{c}{thin film} \\
\cline{4-7}
& & & $\lambda = 1$ & $\lambda = 0$ &  $\lambda = 1$ &  $\lambda = 0$ \\
\hline
 \multicolumn{7}{c}{Diffusion coefficient} \\
Diffusive $l \ll W$ & \multicolumn{1}{r}{$D/v_0 l = $} & $1/2$ & $1/2$ & $1/2$ & $1/3$ & $1/3$\\
Ballistic $W \ll l$ & \multicolumn{1}{r}{$D/v_0 l = $} & $1/2$ & $1/2$ & $(W/\pi l) \ln (l/W)$ & $1/3$ & $(W/4 l) \ln (l/W)$\\
\hline
\multicolumn{7}{c}{Magnetic scattering rate} \\
Diffusive $l \ll W$, weak field $W l \ll l_B^2$ & \multicolumn{1}{r}{$\tau l_B^4/\tau_B W^2 l^2 =$}  & $1/4 \pi^2$ & $1/6$ & $1/6$ & $1/9$ & $1/9$ \\
Ballistic $W \ll l$, weak field $W l \ll l_B^2$ & \multicolumn{1}{r}{$\tau l_B^4/\tau_B W^3 l =$} & $1/4 \pi^3$ & $31 \zeta(5)/\pi^5$ & $1/4\pi$ & $31 \zeta(5)/4 \pi^4$ & $1/16$\\
Ballistic $W \ll l$, strong field $l_B^2 \ll W l$\quad\strut & \multicolumn{1}{r}{$\tau l_B^2/ \tau_B W^2 =$}  & $4/\pi^4$ & $5/24$ & $1/3$ & $5/24$ & $1/3$ \\
\hline\hline
\end{tabular}
\caption{Summary of the results for the diffusion constant and magnetic scattering rate in samples of different geometries and in different regimes.}
\label{tsumm}
\end{table*}

We have studied magnetotransport in disordered metallic samples with restricted quasi-1D or quasi-2D geometry. When the electron mean free path $l$ becomes comparable or less than the system width $W$, the result crucially depends on the type of the sample boundary. We have considered the most general boundary condition, where an electron scattering is either specular or random with the probabilities $\lambda$ and $1 - \lambda$, respectively. While the conductivity of the material is given by the Drude formula $\sigma_0 = e^2\nu D$ in terms of the diffusion constant $D$, its magnetic field dependence also involves the magnetic scattering time $\tau_B$. In the case of a quasi-1D sample in transverse field, magnetoconductivity is given by Eq.\ (\ref{MR}). For the quasi-2D sample in parallel field, an alternative expression (\ref{Ds2D}) applies. Limiting values of the two relevant parameters, $D$ and $\tau_B$, are summarized in Table \ref{tsumm} for different geometries, boundary conditions, and strength of magnetic field.

The diffusion constant of a narrow sample ($W \ll l$) depends on the type of the edge via the parameter $b$ defined in Eq.\ (\ref{b}). The crossover between the mirror and diffuse boundary limits occurs when $b \sim 1$, i.e. $1 - \lambda \sim W/l$ and is described by the functions (\ref{Dphi}) and (\ref{Db2D}) in the quasi-1D and 2D cases, respectively. Note that a small randomness of the boundary scattering may be sufficient to limit the electron mobility in a narrow sample.

Let us note that the electron mean free path, determined experimentally from the mobility measurement, is different from the parameter $l$ of our theory in the case of ballistic sample $W \ll l$. The former quantifies transport scattering rate both on the bulk and boundary imperfections while the latter is related to bulk scattering only. If the measured diffusion constant is interpreted with the use of the classical expression $D = v_0 l_\text{exp}/2$ in a ballistic sample ($W \ll l$) with a boundary roughness $\lambda \sim 1$, it will yield $l_\text{exp} \sim W \ln(l/W)$. This value of $l_\text{exp}$ differs from $W$ only by a logarithmic factor. In many experiments,\cite{Katine, Reulet, Hansen, Schaepers, Kallaher, Chiu, Lin, Muehlbauer} the observed mean free path is of the order of the sample width suggesting the ballistic limit of electron transport limited by the surface roughness.

The magnetic scattering rate $1/\tau_B$ has three limiting values depending on the sample geometry and the strength of the magnetic field: (i) diffusive limit ($l \ll W \ll l_B$), (ii) ballistic weak field limit ($W \ll l$ and $Wl \ll l_B^2$), and (iii) ballistic strong field limit ($W \ll l$ and $l_B^2 \ll Wl$). The crossover between diffusive and ballistic limit in a weak field is governed by the ratio $x = W/l$, see Eq.\ (\ref{tby}). The function $g(x)$ is given by Eqs. (\ref{gC}), (\ref{gMD}), and (\ref{g2D}) for a cylinder, flat quasi-1D channel, and thin quasi-2D film, respectively. In the case of a ballistic sample $W \ll l$, the function $g(x)$ also depends on the boundary scattering parameter $\lambda$. For quasi-1D samples, the corresponding functions are shown in Figs.\ \ref{gCf} and \ref{gMixedF}.

In the limit of a relatively strong magnetic field $W \ll l_B \ll \sqrt{W l}$, magnetic scattering rate is sensitive to the boundary type. The crossover between mirror and diffuse boundary scattering is governed by the parameter $(1 - \lambda) Wl/l_B^2$, see Eqs.\ (\ref{tbStrong}) and (\ref{tbStrong2D}). The corresponding crossover function $F(x)$ is given by Eq.\ (\ref{FF}), see Fig.\ \ref{Fx}. Remarkably, in the same narrow sample ($W \ll l$) electron mobility can be dominated by the surface scattering while magnetoconductance exhibits the dependence characteristic for specular boundaries. This happens when $W/l \ll 1 - \lambda \ll l_B^2/Wl \ll 1$ and suggests a resolution for the discrepancy observed in some previous works, \cite{Katine, Niimi} where the behavior of the diffusion coefficient was better described by the rough boundary model while the magnetoconductance conforms better to mirror boundaries.

To summarize, the diffusion constant and magnetic scattering rate exhibit qualitatively different dependence on the quality of the sample boundary. The underlying reason is that the two quantities are dominated by different trajectories. In a narrow sample, the diffusion constant is determined by trajectories that experience a large number $l/W \gg 1$ of boundary scatterings between two bulk scatterings. This results in a high sensitivity of the diffusion constant to the boundary quality. Already at very small values of the parameter $1 - \lambda \sim W/l \ll 1$, a crossover from mirror to diffuse boundary limit occurs. On the contrary, the magnetic scattering rate is governed by the contribution of shallow trajectories. In the weak field limit, only a few boundary scatterings occur between two consecutive bulk scatterings, hence, the crossover between specular and diffuse boundary occurs at $\lambda \sim 1/2$. In the strong field limit, the dominating trajectories scatter $Wl/l_B^2 \gg 1$ times at the boundary leading again to high sensitivity of the magnetic scattering rate to the boundary roughness. The crossover in this case happens when $1 - \lambda \sim l_B^2/Wl \ll 1$.

% The underlying reason for the contrasting dependence of the diffusion constant, weak field and strong field magnetic scattering rate on the specularity parameter $\lambda$ in the ballistic limit $W\ll l$ is related to the number of boundary scattering experienced by trajectories which provide the main contribution. The diffusion constant is governed by trajectories scattering $l/W \gg 1$ times, resulting in a strong sensitivity to boundary roughness, which manifests itself as a mirror-diffuse crossover at very small values of the boundary roughness parameter $1 - \lambda \sim W/l \ll 1$. Magnetic scattering rate in weak field, on the other hand, is governed by trajectories that scatter a few times at the boundary leading to a smooth dependence on boundary roughness (cf. Fig. \ref{cf}) and a mirror-diffuse crossover at $1 - \lambda \sim 1/2$. Strong field magnetic scattering rate is dominated by trajectories scattering $Wl/l_B^2 \gg 1$ times at the boundary, leading again to strong dependence on boundary roughness and a mirror-diffuse crossover at small values of the boundary roughness parameter $1 - \lambda \sim l_B^2/Wl \ll 1$.

Our results apply to a variety of semiconductor nanostructures with or without spin-orbit coupling, including carbon nanotubes and edge states of topological insulators. For a system with a negligible spin-orbit coupling, one should take the value $\eta = -2$ in Eqs.\ (\ref{MR}) and (\ref{Ds2D}). This corresponds to the conventional weak localization. In the presence of strong spin-orbit coupling, the parameter $\eta = 1$ corresponds to weak antilocalization.\cite{Hikami}. For multiwall carbon nanotubes, where intervalley scattering is important,\cite{McCann} the parameter is $\eta = -2$. The edge states of topological insulators\cite{Lu, Garate} belong to the strong spin-orbit interaction limit with $\eta = 1$, but the overall result should be doubled to account for the two conducting edge channels. This, however, applies only if the coupling between the edges is negligible. In some recent experiments, \cite{Lin} the value of the degeneracy prefactor was observed to be between 1 and 2 showing some finite coupling between the two conducting edges.

\acknowledgments

We are grateful to Igor Gornyi, Yongqing Li, and Alexander Mirlin for helpful and stimulating discussions. The work was supported by the Russian Science Foundation (Grant No. 14-42-00044).

\appendix

\section{Calculation of the Cooperon}
\label{App:Cooperon}

In this Appendix, we find an explicit expression for the reduced Cooperon propagator $\mathcal{R}$ defined in Eq.\ (\ref{R}). We first present the solution for the mixed boundary conditions Eq.\ (\ref{mixed}). The limits of mirror and diffuse boundaries are then deduced from this general result. In the end of this Appendix, we give the expression for the reduced Cooperon for a cylinder with periodic boundary conditions.

The reduced Boltzmann kinetic equation can be written explicitly as
\begin{equation}
\label{RBoltzmann}
 \left[
   1 + l \sin\phi \frac{\partial}{\partial y}
 \right] \mathcal{R}(y,y',\phi,\phi')
  = 2\pi W \delta(\phi - \phi') \delta(y - y').
\end{equation}
Note that we have normalized the unity operator in the right-hand side by the width of the sample. The spatial and angular dependence of the reduced Cooperon can be separated in the following way:
\begin{equation}
 \mathcal{R}(y,y',\phi,\phi')
  = e^{-\tfrac{y}{l \sin\phi} + \tfrac{y'}{l \sin\phi'}} \begin{cases}
      M_+(\phi, \phi'), & y > y', \\
      M_-(\phi, \phi'), & y < y'.
    \end{cases}
 \label{Ry}
\end{equation}
Integrating Eq.\ (\ref{RBoltzmann}) over an infinitesimal interval around $y = y'$, we obtain a relation 
\begin{equation}
\label{Mpm}
 M_+(\phi,\phi') - M_-(\phi,\phi')
  = \frac{2\pi W}{l \sin\phi} \delta(\phi-\phi').
\end{equation}
It is convenient to introduce an auxiliary matrix structure to discriminate between states with positive and negative angle $\phi$. We will write
\begin{equation}
 \hat{\mathcal{R}}
  = \begin{pmatrix}
      \mathcal{R}(\phi, \phi') & \mathcal{R}(\phi, -\phi') \\
      \mathcal{R}(-\phi, \phi') & \mathcal{R}(-\phi, -\phi')
    \end{pmatrix}
 \label{hat}
\end{equation}
and assume both $\phi$ and $\phi'$ positive. The same matrix notations also apply to $\hat M_\pm$. We also introduce the Dirac notations $|a\}$ for functions $a(\phi)$ of the positive angle $\phi$ with the following weighted scalar product:
\begin{equation}
 \{ a | b \}
  = \frac{1}{2} \int_0^\pi d\phi\, \sin\phi\, a(\phi) b(\phi).
 \label{sinmeasure}
\end{equation}
Such notations are particularly suitable for describing the boundary conditions. With these definitions, the identity (\ref{Mpm}) can be rewritten as
\begin{equation}
 \hat M_+ - \hat M_-
  = \frac{\pi W}{l} \begin{pmatrix}
      1 & 0 \\
      0 & -1
    \end{pmatrix}.
 \label{Mpm2}
\end{equation}
The unity operators in the right-hand side absorb the denominator $\sin\phi$ of Eq.\ (\ref{Mpm}) due to the integration measure (\ref{sinmeasure}) in the $\phi$ space.

The reduced propagator $\mathcal{R}$ obeys the same boundary conditions as the full Cooperon $\mathcal{C}$. With the Dirac notations introduced above, we can rewrite the boundary conditions in the operator form
\begin{gather}
 \mathcal{R}(y = \pm W/2, \mp \phi)
  = K \mathcal{R}(y = \pm W/2, \pm \phi), \label{Rboundary} \\
 K
  = \lambda + (1 - \lambda) | 1 \} \{ 1 |. \label{K}
\end{gather}
The operator $K$ corresponds to the general mixed boundary Eq.\ (\ref{mixed}) with the probabilities $\lambda$ and $1 - \lambda$ of mirror and diffuse scattering, respectively. The representation (\ref{K}) demonastrates another advantage of the Dirac notations (\ref{sinmeasure}): the diffuse scattering has the form of a simple projector on to the constant function $| 1 \}$.

Using Eqs.\ (\ref{Ry}) and (\ref{hat}), we can formulate the boundary conditions (\ref{Rboundary}) in terms of $\hat M_\pm$,
\begin{gather}
 \begin{pmatrix}
  -\tilde K, & 1
 \end{pmatrix} \hat M_+
  = 0,
 \qquad
 \begin{pmatrix}
  1, & -\tilde K
 \end{pmatrix} \hat M_-
  = 0, \label{Mboundary} \\
 \tilde K
  = h K h, 
 \qquad\qquad
 h
  = e^{-\tfrac{W}{2l \sin\phi}}. \label{Kh}
\end{gather}
Together with Eq.\ (\ref{Mpm2}), this fully determines the matrices $M_\pm$. Multiplying Eq.\ (\ref{Mpm2}) by the row $(1, - \tilde K)$ from the left, we cancel $M_-$ and obtain an equation involving $M_+$ only. This equation can be combined with the first identity of Eq.\ (\ref{Mboundary}) into a single matrix condition
\begin{equation}
 \begin{pmatrix}
  1 & -\tilde K \\
  -\tilde K & 1
 \end{pmatrix} M_+
  = \frac{\pi W}{l} \begin{pmatrix}
      1 & \tilde K \\
      0 & 0
    \end{pmatrix}.
\end{equation}
The solution for $M_+$ is now given by the matrix operator inverse. Similar manipulations lead to the matrix equation for $M_-$. Explicitly, the solution for $\hat M_\pm$ can be written as
\begin{align}
 \hat M_+
  &= \frac{\pi W}{l} \begin{pmatrix}
      1 \\ \tilde K
    \end{pmatrix} \bigl( 1 - \tilde K^2 \bigr)^{-1} \begin{pmatrix}
      1, & \tilde K
    \end{pmatrix}, \label{MpK} \\
 \hat M_-
  &= \frac{\pi W}{l} \begin{pmatrix}
      \tilde K \\ 1
    \end{pmatrix} \bigl( 1 - \tilde K^2 \bigr)^{-1} \begin{pmatrix}
      \tilde K, & 1
    \end{pmatrix}. \label{MmK}
\end{align}
Thus we have reduced the computation of the Cooperon to the calculation of the operator inverse
\begin{equation}
 (1 - \tilde K^2)^{-1}
  = \frac{(1 - \tilde K)^{-1} + (1 + \tilde K)^{-1}}{2}.
\end{equation}
Using the Dirac notations (\ref{sinmeasure}) and the explicit form of the boundary operator Eq.\ (\ref{K}), we have the following expression for the two terms entering the above equation:
\begin{gather}
 (1 \pm \tilde K)^{-1}
  = \frac{1}{1 \pm \lambda h^2} \mp \frac{h | 1 \}}{1 \pm \lambda h^2} \frac{1 - \lambda}{\Gamma_\pm} \frac{\{ 1 | h}{1 \pm \lambda h^2}, \\
 \Gamma_\pm
  = \{ 1 | \frac{1 \pm h^2}{1 \pm \lambda h^2} | 1 \}
  = \frac{1}{2} \int_0^\pi d\phi\, \sin\phi\, \frac{1 \pm h^2}{1 \pm \lambda h^2}.
\end{gather}
Upon substitution of this result into Eqs.\ (\ref{MpK}) and (\ref{MmK}) and then into Eq.\ (\ref{Ry}), we obtain the following general expression for the reduced Cooperon propagator with the mixed boundary conditions:
\begin{widetext}
\begin{multline}
 \label{Rf}
 \mathcal{R}(y,y',\phi,\phi')
  = e^{-\tfrac{y}{l \sin\phi} + \tfrac{y'}{l \sin\phi'}} \frac{W}{l} \Biggl\{ \frac{\pi}{2}
      (1 - \lambda)\left[
        \frac{h h'}{(1 - \lambda h^2) \Gamma_- (1 - \lambda {h'}^2)}
        -\frac{\mathop{\mathrm{sign}}(\phi \phi') h h'}{(1 + \lambda h^2) \Gamma_+ (1 + \lambda {h'}^2)}
      \right] \\
      +\frac{2\pi \lambda h^2 \bigl[ \lambda h^2 \delta(\phi - \phi') + \delta(\phi + \phi') \bigr]}{(1 - \lambda^2 h^4) |\sin\phi|}
      +\frac{2 \pi \delta(\phi - \phi')}{|\sin\phi|} \theta[(y - y') \phi]
    \Biggr\}.
\end{multline}
\end{widetext}
Here we have also unfolded the matrix structure of Eq.\ (\ref{hat}). The function $h$ has an implicit argument $\phi$ as defined in Eq.\ (\ref{Kh}) while $h' = h(\phi')$.

The three terms in curly braces in Eq.\ (\ref{Rf}) represent different trajectories contributing to the Cooperon. The first term is proportional to $1 - \lambda$ and corresponds to trajectories that experience diffuse scattering at the boundary at least once. The second term $\sim \lambda$ describes trajectories scattering only specularly hence initial and final angles are perfectly correlated, $\phi = \pm \phi'$. The third term is independent of $\lambda$ and represents the straight trajectory propagating directly from the initial to the final point.

In the case of cylinder geometry the calculation of the Cooperon is much simpler. Equation (\ref{Mpm}) is supplemented by the periodic boundary condition that has the form $h^2 M_+ = M_-$. Solving the two equations for $M_\pm$ and using Eq.\ (\ref{Ry}) we obtain the result
\begin{equation}
 \label{Rc}
 \mathcal{R}(y,y',\phi,\phi')
  = e^{-\tfrac{y - y'}{l \sin\phi}}\;
    \frac{\pi W \delta(\phi - \phi')\, e^{\tfrac{\mathop{\mathrm{sign}}(y - y') W}{2l \sin\phi}}}{l \sin \phi \sinh \bigl( \tfrac{W}{2 l \sin\phi} \bigr)}.
\end{equation}
Let us recall that $W$ is the circumference of the cylinder in this case.

\section{Diagrammatic derivation of Eq.\ (\protect\ref{Ds})}
\label{App:diag}

In this Appendix, we outline the derivation of the weak localization correction to the conductivity, Eq.\ (\ref{Ds}), for a ballistic system with mixed boundary conditions using the diagrammatic approach.

\subsection{Classical conductivity}

\begin{figure}
 \includegraphics[width=\columnwidth]{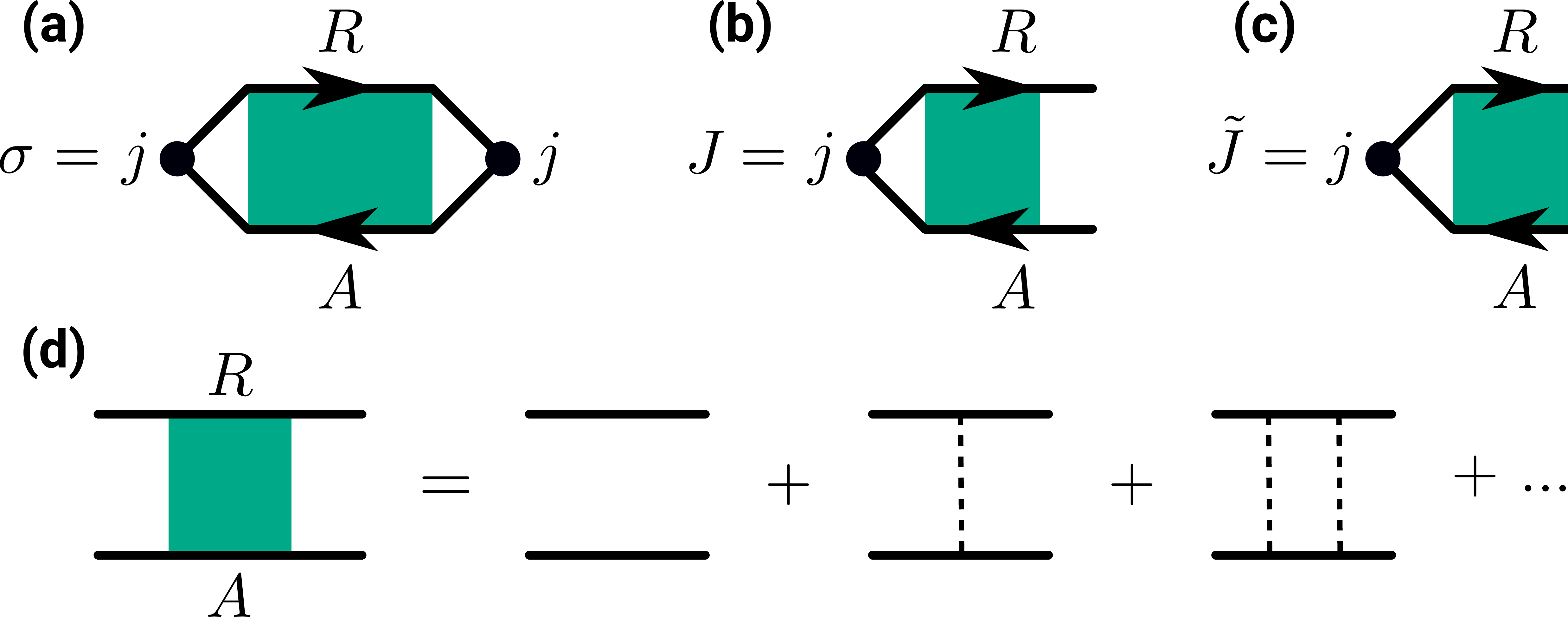}
 \caption{(a) Diagram for the Drude conductivity involving two current operators $j$, Eq.\ (\protect\ref{currents}), and a diffuson propagator (d), Eq.\ (\protect\ref{BS}). Fully (b) and partially (c) dressed current operators, Eqs.\ (\protect\ref{currents}) and (\protect\ref{tJ}). }
 \label{fig:drude}
\end{figure}

Conductivity of a disordered system is given by the Kubo formula in terms of the average product of two Green functions and two current operators. The classical (Drude) result implies averaging the two Green functions independently. When impurity scattering is anisotropic, the vertex correction should be included to account for the difference between quantum and transport scattering rates. Diagrammatic representation of the Drude conductivity is shown in Fig.\ \ref{fig:drude}a. Solid lines in the diagrams represent disorder-averaged Green functions
\begin{equation}
 G^{R,A}_\mathbf{p} = \bigl[ \epsilon - \xi(\mathbf{p}) \pm i/2\tau \bigr]^{-1}
\end{equation}
at the Fermi energy $\epsilon$ with $\xi(\mathbf{p})$ being the underlying electron dispersion. Dashed lines in the diagrams stand for the disorder correlation functions. We will include the boundary scattering in the definition of the dashed lines below.

The ladder diagrams of the vertex correction sum up to the diffuson propagator, Fig.\ \ref{fig:drude}d. If the system possesses time-reversal symmetry (in the absence of magnetic field) diffuson and Cooperon propagators coincide. An individual ladder rung is given by the product of two Green functions integrated over the absolute value of momentum. Using the quasiclassical approximation, integration over $|p|$ can be replaced by the integration over $\xi$ yielding (throughout this Appendix we assume $\hbar = 1$)
\begin{equation}
 P_\mathbf{q}(\mathbf{n})
  = \int \frac{d\xi}{2\pi}\, G^R_{\mathbf{p} + \mathbf{q}} G^A_{\pm\mathbf{p}}
  = \left[ \frac{1}{\tau} + i v_0 \mathbf{n} \mathbf{q} \right]^{-1}.
\end{equation}
Here, $\mathbf{n}$ is the unit vector in the direction of $\mathbf{p}$. The sign of momentum in $G^A(\pm\mathbf{p})$ distinguishes between diffuson and Cooperon but the value of $P(\mathbf{n}, \mathbf{q})$ is manifestly independent of this sign. We will regard $P$ as an operator acting in position and velocity space similar to the treatment of the Cooperon in the main text of the paper. Dependence on $\mathbf{q}$ defines the real space structure of $P$ while $\mathbf{n}$ corresponds to the velocity direction on the Fermi surface. Thus $P$ is diagonal in the velocity space.

The diffuson propagator is given by the sum of the ladder diagrams in Fig.\ \ref{fig:drude}d. Using the operator notations, we can represent the corresponding Bethe-Salpeter equation as
\begin{gather}
 \mathcal{D}
  = P + P \mathcal{V D},
 \qquad
 \mathcal{V}(\mathbf{r}; \mathbf{n}, \mathbf{n}')
  = \frac{1}{\tau} + \mathcal{V}_b(\mathbf{r}; \mathbf{n}, \mathbf{n}').
 \label{BS}
\end{gather}
Here, the vertex part $\mathcal{V}$ is an operator in the same position-velocity space as $P$. The operator $\mathcal{V}$ is diagonal in the real space and contains the isotropic bulk scattering term $1/\tau$ and the boundary scattering term $\mathcal{V}_b$. The latter is introduced to impose the boundary conditions and is effective only near the sample edges. For a flat 2D sample, we have
\begin{multline}
 \mathcal{V}_b(y; \phi, \phi')
  = \delta(y - W/2 - 0) \mathcal{V}_+(\phi, \phi') \\ + \delta(y + W/2 + 0) \mathcal{V}_-(\phi, \phi').
 \label{Vb}
\end{multline}
Here the angles $\phi$ and $\phi'$ denote the directions of $\mathbf{n}$ and $\mathbf{n}'$. The angular parts of the vertex $\mathcal{V}_\pm$ encode the scattering properties of the upper and lower boundaries of the sample.

To demonstrate the emergence of the boundary conditions from the vertex (\ref{Vb}), we convert the operator $P$ to the real-space representation in the transverse $y$ direction,
\begin{equation}
 P(y, y', q; \phi)
  = e^{-\tfrac{y - y'}{l \sin\phi}(1 + i q l \cos\phi)}\; \frac{\theta \bigl[ (y - y') \phi \bigr]}{v_0 |\sin\phi|}.
\end{equation}
The conserved $x$ component of the momentum is denoted by $q$. Upon substitution of the above expression into the Bethe-Salpeter Eq.\ (\ref{BS}) and taking $y = \pm W/2$ for $\phi \lessgtr 0$, we obtain
\begin{equation}
 \mathcal{D}(\pm W/2, \phi)
  = \int \frac{d\phi'}{2\pi v_0 |\sin\phi'|} \mathcal{V}_\pm(\phi, \phi') \mathcal{D}(\pm W/2, \phi').
\end{equation}
In this expression we suppress the arguments $q$, $y'$ and $\phi'$ of the propagators. The mixed boundary conditions (\ref{mixed}) are reproduced with the scattering vertex of the following form:
\begin{multline}
 \mathcal{V}_+(\phi, \phi')
  = \mathcal{V}_-(\phi, \phi') \\
  = \pi v_0 |\sin\phi| \left[ 2\lambda\delta(\phi + \phi') + (1 - \lambda) |\sin\phi'| \theta(-\phi \phi') \right].
\end{multline}

In the absence of magnetic field, the diffuson propagator $\mathcal{D}$ coincides with the Cooperon and differs by a factor of $\tau$ from the reduced propagator (\ref{R}) when the longitudinal momentum is $q = 0$. The classical Drude conductivity is given by the diagram in Fig.\ \ref{fig:drude}a involving one bare and one dressed current operator, Fig.\ \ref{fig:drude}b:
\begin{equation}
 j_x
  = e v_0 n_x
  = e v_0 \cos\phi,
 \qquad
 J_x
  = \mathcal{D} j_x.
 \label{currents}
\end{equation}
With the diffuson propagator defined above, this yields
\begin{equation}
 \sigma_0
  = e^2 \nu D
  = \nu \langle\!\langle j_x J_x \rangle\!\rangle
  = \nu \langle\!\langle j_x \mathcal{D} j_x \rangle\!\rangle.
 \label{sigma0}
\end{equation}
Let us recall that the double angular brackets $\langle\!\langle \ldots \rangle\!\rangle$ denote averaging with respect to both position and direction, see Eq.\ (\ref{avC}). The last result is equivalent to the expression for the diffusion constant (\ref{D}) in terms of the reduced propagator (\ref{R}).

\subsection{Quantum correction}

\begin{figure}
 \includegraphics[width=\columnwidth]{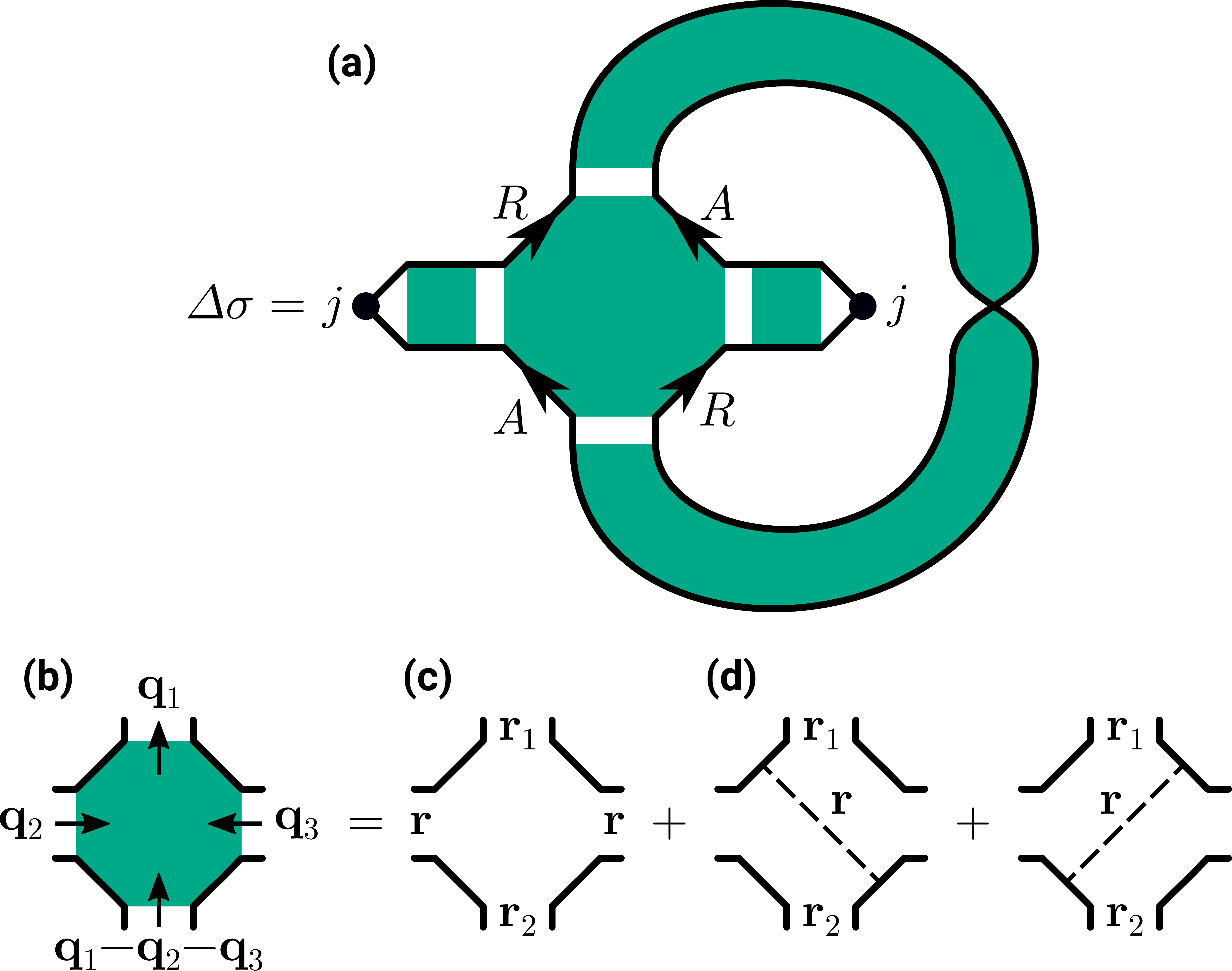}
 \caption{(a) Diagram for the quantum correction to the conductivity. The Hikami box (b) contains one empty (c) and two equal crossed (d) diagrams. Momentum labels in (b) correspond to Eqs.\ (\protect\ref{GGGG}) and (\protect\ref{GGG}). Real space labels in (c) and (d) are used in Eqs.\ (\protect\ref{H1}) and (\protect\ref{H2}).}
 \label{fig:hikami}
\end{figure}

Interference correction to the conductivity is due to the set of maximally crossed diagrams\cite{Gorkov}, shown in Fig.\ \ref{fig:hikami}a. These diagrams involve one loop of the Cooperon propagator closed by the Hikami box. We outline the calculation of the Hikami box for an arbitrary boundary conditions and geometry of the sample.

An ``empty'' Hikami box, Fig.\ \ref{fig:hikami}c, involves an integral of four Green functions. This integral can be readily written in momentum representation (notations are explained in Fig.\ \ref{fig:hikami}a)
\begin{multline}
 \int \frac{d\xi}{2\pi}\, G^R_\mathbf{p} G^A_{-\mathbf{p} + \mathbf{q}_1} G^R_{-\mathbf{p} + \mathbf{q}_1 - \mathbf{q}_3} G^A_{\mathbf{p} - \mathbf{q}_2} \\
  = P_{\mathbf{q}_1}(\mathbf{n}) P_{\mathbf{q}_1 - \mathbf{q}_2 - \mathbf{q}_3}(-\mathbf{n}) \bigl[ P_{\mathbf{q}_2}(\mathbf{n}) + P_{\mathbf{q}_3}(\mathbf{n}) \bigr].
 \label{GGGG}
\end{multline}
The current vertices attached to the Hikami box are either bare (if no disorder scattering lines are inserted) or dressed by diffuson ladders that end with the impurity scattering see Fig.\ \ref{fig:drude}c. We will denote such currents as
\begin{equation}
 \tilde J
  = j + j \mathcal{D V}
  = j + J \mathcal{V}
  = J P^{-1}.
 \label{tJ}
\end{equation}

The $P$ operators in Eq.\ (\ref{GGGG}) complement one of the two current vertices attached to the Hikami box to form the fully dressed current $J$. We convert Eq.\ (\ref{GGGG}) into the real space representation (see labels in Fig.\ \ref{fig:hikami}c) and obtain
\begin{equation}
 H_1
  = -2\int d\mathbf{r}\; P_{\mathbf{r}_2, \mathbf{r}}(-\mathbf{n}) P_{\mathbf{r}, \mathbf{r}_1}(\mathbf{n}) \tilde J_\mathbf{r}(\mathbf{n}) J_\mathbf{r}(\mathbf{n}).
 \label{H1}
\end{equation}
The two remaining $P$ operators in this expression represent the initial and final step of the Cooperon loop, see Fig.\ \ref{fig:hikami}a. The quantum correction due to the ``empty'' Hikami box is thus
\begin{equation}
 \Delta\sigma_1
  = -\frac{1}{\pi} \int d\mathbf{r}\, d\mathbf{n}\;
      \tilde J_\mathbf{r}(\mathbf{n}) J_\mathbf{r}(\mathbf{n})\; \mathcal{C}_{\mathbf{r}, \mathbf{r}}(\mathbf{n}, -\mathbf{n}).
 \label{Dsigma1}
\end{equation}
Here the Cooperon propagator gives the probability of return to the same point with the opposite velocity direction. Therefore $\Delta\sigma_1$ represents the backscattering interference correction.\cite{Dmitriev}

The crossed Hikami box is presented in Fig.\ \ref{fig:hikami}d. The two ways of crossing the square yield identical contributions, therefore we compute the first of the two crossed diagrams and double the result. The crossed box contains two momentum integrals, each involving three Green functions, and one disorder scattering vertex $\mathcal{V}$. Similar to the case of the empty box discussed above, we first perform integration in the momentum space. The directions of the two momenta are labeled by $\mathbf{n}$ and $\mathbf{n}'$
\begin{subequations}
\label{GGG}
\begin{align}
 &\int \frac{d\xi}{2\pi}\, G^R_\mathbf{p} G^A_{-\mathbf{p} + \mathbf{q}_1} G^R_{-\mathbf{p} + \mathbf{q}_1 - \mathbf{q}_3} \notag \\
 & \hspace{4cm} = -i P_{\mathbf{q}_1} (\mathbf{n}) P_{\mathbf{q}_3} (\mathbf{n}), \\
 &\int \frac{d\xi'}{2\pi}\, G^R_{-\mathbf{p}' + \mathbf{q}_1 - \mathbf{q}_3} G^A_{\mathbf{p}' - \mathbf{q}_2} G^R_{\mathbf{p}'} \notag \\
 & \hspace{2.5cm} = -i P_{\mathbf{q}_1 - \mathbf{q}_2 - \mathbf{q}_3} (-\mathbf{n}') P_{\mathbf{q}_2} (\mathbf{n}').
\end{align}
\end{subequations}
Similar to the case of the empty box, the second $P$ operators in Eqs.\ (\ref{GGG}) complement the current vertices $\tilde J$ to form the two fully dressed currents $J$. We convert the above integrals together with currents to the real space representation and insert the single disorder scattering term. This yields the crossed Hikami box in the form
\begin{equation}
 H_2
  = 2\int d\mathbf{r}\; P_{\mathbf{r}_2, \mathbf{r}}(-\mathbf{n}') P_{\mathbf{r}, \mathbf{r}_1}(\mathbf{n})
    J_\mathbf{r}(\mathbf{n}') \mathcal{V}_\mathbf{r}(\mathbf{n}', \mathbf{n}) J_\mathbf{r}(\mathbf{n}).
 \label{H2}
\end{equation}
The two $P$ operators in this expression are the first and the last step of the Cooperon loop. The quantum correction to the conductivity with the crossed box is
\begin{equation}
 \Delta\sigma_2
  = \frac{1}{\pi} \int d\mathbf{r}\, d\mathbf{n}\, d\mathbf{n'}\;
      J_\mathbf{r}(\mathbf{n}') \mathcal{V}_\mathbf{r}(\mathbf{n}', \mathbf{n}) J_\mathbf{r}(\mathbf{n})\;
      \mathcal{C}_{\mathbf{r}, \mathbf{r}}(\mathbf{n}, -\mathbf{n}').
 \label{Dsigma2}
\end{equation}
This term involves the Cooperon propagator with two different velocity arguments and represents the nonbackscattering correction.\cite{Dmitriev}

In order to combine the two parts of the quantum correction $\Delta\sigma_{1,2}$, we recall that the leading contribution to the Cooperon at coincident points is due to the soft mode discussed in detail in Sec.\ \ref{Sec:general}. This soft mode is isotropic in the velocity space and constant in real space. Hence we can replace the Cooperons in Eqs.\ (\ref{Dsigma1}) and (\ref{Dsigma2}) by their angular and spatial averages. This leads to
\begin{equation}
 \Delta\sigma
  = -\frac{\langle\mathcal{C}_{\mathbf{r}, \mathbf{r}}\rangle}{\pi}
    \int d\mathbf{r}\, d\mathbf{n}\; [\tilde J-J\mathcal{V}]_\mathbf{r}(\mathbf{n}) J_\mathbf{r}(\mathbf{n}).
 \label{DsigmaAll}
\end{equation}
According to Eq.\ (\ref{tJ}), the combination $\tilde J - J \mathcal{V}$ equals the bare current operator $j$ and the integral in Eq.\ (\ref{DsigmaAll}) reproduces the classical Drude conductivity $\sigma_0$, see Eq.\ (\ref{sigma0}). This way we have derived the quantum correction (\ref{Ds}) with $\eta = -1$. Extra spin degeneracy doubles the above result leading to $\eta = -2$.

In a sample with strong spin-orbit coupling, the Cooperon propagator becomes a non-trivial operator in the spin space. The soft mode that determines the quantum correction has a structure of the spin singlet.\cite{Akkermans} Boundary scattering acts in a complicated way on the other three triplet modes but does not mix them with the singlet.\cite{Kettemann, Wenk} The derivation outlined in this Appendix can be generalized to account for the singlet Cooperon channel in the case of strong spin-orbit coupling. The quantum correction has the same form of Eq.\ (\ref{DsigmaAll}) but with an opposite sign (weak antilocalization). This corresponds to the case $\eta = +1$ of Eq.\ (\ref{Ds}).

If the system possesses some extra degeneracy, the factor $\eta$ in Eq.\ (\ref{Ds}) should be further modified to take it into account. For instance, in a thick metallic carbon nanotube, electrons obey linear dispersion law with a pseudospin degree of freedom strongly coupled to the momentum. This leads to the positive quantum correction to the conductance. An extra valley and spin degeneracy yield $\eta = +4$.

\bibliography{refs}

\end{document}